\documentclass[10pt]{article}
\usepackage[T1]{fontenc}
\usepackage[latin9]{inputenc}

\makeatletter

\providecommand{\tabularnewline}{\\}

\usepackage{enumitem}		

\usepackage{M12}

\numberwithin{equation}{section}

\AtBeginDocument{
  
}

\makeatother

\begin{document}
\setpreprint{TIFR/TH/12-35, \href{http://arxiv.org/abs/1211.5587}{arXiv:1211.5587}}

\title{Comment on Strings in \texorpdfstring{$AdS_{3}\times S^{3}\times S^{3}\times S^{1}$}{AdS3 x S3 x S3 x S1} at One Loop}

\author{Michael C. Abbott \address{Tata Institute of Fundamental Research,\\
Homi Bhabha Rd, Mumbai 400-005, India.\\
abbott@theory.tifr.res.in} }

\date{23 November 2012}
\maketitle
\begin{abstract}
This paper studies semiclassical strings in $AdS_{3}\times S^{3}\times S^{3}\times S^{1}$
using the algebraic curve. Calculating one-loop corrections to the
energy of the giant magnon fixes the constant term $c$ in the expansion
of the coupling $h(\lambda)$. Comparing these to similar corrections
for long spinning strings gives a prediction for the one-loop term
$f_{1}$ in the expansion of the cusp anomalous dimension $f(h)$,
for all $\alpha$ (where $\alpha\to1$ is the $AdS_{3}\times S^{3}\times T^{4}$
limit). For these semiclassical mode sums there is a similar choice
of regularisation prescriptions to that encountered in $AdS_{4}\times CP^{3}$.
However at $\alpha\neq\tfrac{1}{2}$ they lead to different values
of $f_{1}$ and are therefore not related by a simple change of the
coupling. The algebraic curve is also used to calculate various finite-size
corrections for giant magnons, which are well-behaved as $\alpha\to1$,
and can be compared to the recently published S-matrices. 
\end{abstract}

\section{Introduction}

The usual starting point for discussing integrable strings in $AdS_{5}\times S^{5}$
is the Metsaev--Tseytlin coset action \cite{Metsaev:1998it,Arutyunov:2009ga},
where classical integrability follows from the fact that the coset
is a Riemannian symmetric space \cite{Mandal:2002fs,Bena:2003wd}.
This is the strong-coupling end of the best-studied example of AdS/CFT,
and the integrable structure now extends to all values of the 't Hooft
coupling $\lambda$ \cite{Beisert:2010jr}. The same statements are
true for the second-best-studied example, with strings on $AdS_{4}\times CP^{3}$
\cite{Aharony:2008ug,Arutyunov:2008if,Stefanski:2008ik,Gomis:2008jt,Klose:2010ki}. 

One of the challenges of studying integrability in backgrounds such
as the $AdS_{3}\times S^{3}\times T^{4}$ arising from the D1-D5 system
\cite{Strominger:1996sh,Mandal:2000rp,Adam:2007ws,Sorokin:2011rr}
is the presence of flat directions, and hence massless modes, which
are not captured by the coset action. This is also true of the $AdS_{3}\times S^{3}\times S^{3}\times S^{1}$
background studied here, which has a parameter $\alpha=\cos^{2}\phi$
controlling the relative size of the two 3-spheres \cite{Gauntlett:1998kc},
and hence the masses of the modes in these directions. One of the
reasons this space is interesting is that in the limit $\alpha\to1$
one $S^{3}$ decompactifies to give (when combined with the $S^{1}$)
a $T^{4}$ factor. In this limit two more bosonic modes become massless,
and it is hoped that we may learn about how to handle massless modes
by studying this process. 

There is no known CFT$_{2}$ gauge theory dual for general $\alpha$
\cite{deBoer:1999rh,Gukov:2004ym}, although at $\alpha=1$ there
is a symmetric product-space CFT \cite{Maldacena:1998bw} as well
as more recently a spin chain \cite{Pakman:2009mi} and some work
on magnons \cite{David:2008yk,David:2010yg}. At $\alpha\neq1$ there
is much recent work on integrability \cite{Babichenko:2009dk,Zarembo:2010sg,Zarembo:2010yz,OhlssonSax:2011ms,Forini:2012bb,Rughoonauth:2012qd,Sundin:2012gc,Cagnazzo:2012se,Sax:2012jv}
perhaps the highlight of which is a conjectured all-order Bethe ansatz
for all $\alpha$ \cite{OhlssonSax:2011ms}. This similarly omits
the massless modes (as well as the heavy modes, discussed below) but
has a good $\alpha\to1$ limit. As we have learned from the $AdS_{4}\times CP^{3}$
correspondence, the Bethe equations give the spectrum in terms of
a coupling $h(\mbox{\ensuremath{\lambda}})$ whose relationship to
$\lambda$ (or more precisely here to $R^{2}/\alpha'$) must be found
experimentally \cite{Nishioka:2008gz,Gaiotto:2008cg,Grignani:2008is,Gromov:2008qe,McLoughlin:2008he}.
In this case the strong-coupling expansion is 
\[
h=2g+c+\mathcal{O}\Big(\frac{1}{g}\Big)\qquad\qquad g=\frac{R^{2}}{4\pi\alpha'}=\frac{\sqrt{\lambda}}{4\pi}\gg1.
\]
Here $R$ is the radius of the $AdS_{3}$ part of the spacetime, and
the spheres' $R_{\pm}$ are as follows: 
\begin{equation}
ds^{2}=R^{2}\: ds_{AdS_{3}}^{2}+\frac{R^{2}}{\cos^{2}\phi}\: ds_{S_{+}^{3}}^{2}+\frac{R^{2}}{\sin^{2}\phi}\: ds_{S_{-}^{3}}^{2}+R^{2}\: d\psi^{2}.\label{eq:the-metric}
\end{equation}
The BMN point particle (which is the spin chain vacuum) has momentum
on both of the spheres: the solution is $\vartheta_{+}+\vartheta_{-}=\tau$
in terms of the two azimuthal angles. The two bosonic massless modes
are are fluctuations in $\psi$ and in $\vartheta_{\perp}=\tan\phi\:\vartheta_{+}+\cot\phi\:\vartheta_{-}$.
Both of these are absent from the coset model $D(1,2;\alpha)^{2}/SU(1,1)\times SU(2)^{2}$.
The algebraic curve for this was introduced by Babichenko, Stefa\'{n}ski
and Zarembo in \cite{Babichenko:2009dk}.

The goal of this paper is to use this to calculate (or to guide the
calculation of) semiclassical energy corrections for various classical
string solutions. Such corrections have played an important role in
the past \cite{Frolov:2002av,Frolov:2003tu,Park:2005ji,McLoughlin:2010jw}.
As in $AdS_{4}\times CP^{3}$ there is a distinction between light
modes, which are excitations of the Bethe equations, and heavy modes
which are in some senses composite objects, and because of this there
are similar issues of regularisation \cite{McLoughlin:2008he,McLoughlin:2008ms,Alday:2008ut,Krishnan:2008zs,Gromov:2008fy,Shenderovich:2008bs,Bandres:2009kw,Abbott:2010yb,Astolfi:2011ju,Astolfi:2011bg}.
However (as we will see) in this case this choice \emph{cannot} always
be absorbed into a modification of the coupling constant. 

The two classical systems to be studied are long spinning strings
in $AdS_{3}$, and giant magnons in $S^{3}$. In both cases the classical
solutions are identical to those in $AdS_{5}\times S^{5}$, apart
from momentum on some $S^{1}$ factors. \newcommand{\zz}{\ensuremath{\widetilde{J}}} 
\begin{itemize}
\item Giant magnons have \cite{Hofman:2006xt,OhlssonSax:2011ms} \vspace{-2mm}
\begin{align}
\Delta-J' & =\sqrt{m_{r}^{2}+4h^{2}\sin^{2}\frac{p}{2}}\label{eq:one-magnon-disp}\\
 & =4g\sin\frac{p}{2}+2c\sin\frac{p}{2}+\mathcal{O}\Big(\frac{1}{g}\Big)\nonumber 
\end{align}
where the mass $m_{r}$ depends on which sphere the solution lives
in:
\[
m_{1}=\sin^{2}\phi=1-\alpha\,,\qquad m_{3}=\cos^{2}\phi=\alpha\,.
\]
Using the algebraic curve formalism of \cite{Babichenko:2009dk,Zarembo:2010yz}
to calculate the one-loop correction to the energy $\delta E$ allows
us to find $c$. As in $AdS_{4}\times CP^{3}$ the result depends
on the regularisation used; with a cutoff on the physical energy it
is 
\[
c_{\mathrm{phys}}=\frac{\alpha\log\alpha+(1-\alpha)\log(1-\alpha)}{2\pi}.\tag{\ref{eq:c-magnon}}
\]

\item Long spinning strings have \cite{Gubser:2002tv,Frolov:2006qe} 
\begin{align}
\Delta-S & =f(\lambda)\log S,\qquad f=2h+f_{1}+\smash{\mathcal{O}\Big(\frac{1}{h}\Big)}\label{eq:spinning-Delta-minus-S}\\
 & =4g\log S+\delta\Delta+\mathcal{O}\Big(\frac{1}{g}\Big).\nonumber 
\end{align}
Mode frequencies for these strings were calculated by \cite{Forini:2012bb},
and will be used here to discuss the dependence of the one-loop term
$\delta\Delta$ on the regularisation prescription. Unlike the giant
magnons, the relevant term from integrability $f_{1}$ depends on
the one-loop part of the dressing phase $\sigma_{\mathrm{HL}}$ \cite{Hernandez:2006tk,Beisert:2006ib,Beisert:2006ez,Vieira:2010kb,Casteill:2007ct,Gromov:2008qe}. 
\end{itemize}
Comparing results from these two systems gives a prediction for $f_{1}$
which indicates that the dressing phase must be different to that
seen in $AdS_{5}\times S^{5}$ and $AdS_{4}\times CP^{3}$. This prediction
appears to depend on the regularisation used, but demanding that it
is well-behaved as $\alpha\to1$ rules out the cutoff in the spectral
plane. 

The calculation of $\delta E$ for the giant magnon can also easily
be extended to include finite-size corrections. The L\"{u}scher F-terms
calculated this way can normally be compared term-by-term to certain
(diagonal) elements of the S-matrix, which was until this week unknown. 

However  while this paper was being prepared for publication, two
papers appeared each aiming to derive the S-matrix for $AdS_{3}\times S^{3}\times S^{3}\times S^{1}$
\cite{Ahn:2012hw,Borsato:2012ud}. A preliminary comparison of our
results shows agreement with the elements in both of these, modulo
some issues of phases.

\subsection*{Outline}

Section \ref{sec:Algebraic-Curves} reviews the setup of the algebraic
curve, and section \ref{sec:Defining-cutoff-prescriptions} the various
cutoff prescriptions. Section \ref{sec:Giant-Magnons} uses all of
this for giant magnons. Section \ref{sec:Long-Spinning-Strings} looks
at summing frequencies for the spinning string, and what we can learn
from the comparison. Section \ref{sec:Comments-and-Conclusions} has
a summary and \ref*{enu:the-last-comment} comments. 

Appendix \ref{sec:Finite-Size} looks at finite-size corrections (classical
and one-loop) and the comparison to the proposed S-matrices. Appendix
\ref{sec:Algebraic-Curves-sans-T4} looks briefly at the algebraic
curve for $AdS_{3}\times S^{3}\times T^{4}$, and the matching of
corrections in this limit. Appendix \ref{sec:Worldsheet-Giant-Magnons}
deals with classical giant magnons in the sigma-model.

\section{Algebraic Curves for \texorpdfstring{$AdS_{3}\times S^{3}\times S^{3}$}{AdS3 x S3 x S3}
sans $S^{1}$ \label{sec:Algebraic-Curves}}

The algebraic curve, or finite-gap method, is a way of writing classical
string solutions as Riemann surfaces \cite{Kazakov:2004qf,Kazakov:2004nh,SchaferNameki:2004ik,Beisert:2005bm}.
The Lax connection (which depends on the spectral parameter $x\in\mathbb{C}$)
is integrated around the worldsheet, and the path-ordered exponential
of this is the monodromy matrix, whose eigenvalues are $e^{\pm ip_{\ell}}$
with $p_{\ell}(x)$ called quasimomenta. These contain essentially
all the information about the solution.%
\footnote{But see \cite{Janik:2012ws} for some important caveats. %
} This formalism has been especially useful for semiclassical quantisation,
where vibrational modes are represented by small perturbations of
the quasimomenta \cite{Gromov:2007aq,Vicedo:2008jy,Gromov:2008ec}. 

The setup described in this section is largely from \cite{Babichenko:2009dk},
see also \cite{Zarembo:2010yz}. It starts from the Cartan matrix
for $d(2,1;\alpha)^{2}$. Since this is a continuous family of distinct
Lie super-algebras, $A_{\ell m}$ has non-integer entries: 
\begin{equation}
A=\left[\begin{array}{ccc}
4\sin^{2}\phi & -2\sin^{2}\phi & 0\\
-2\sin^{2}\phi & 0 & -2\cos^{2}\phi\\
0 & -2\cos^{2}\phi & 4\cos^{2}\phi
\end{array}\right]\otimes1_{2\times2}\,.\label{eq:cartan-A}
\end{equation}
For each Cartan generator $\Lambda_{\ell}$ there is a quasimomentum
$p_{\ell}(x)$, where $\ell=1,2,3,\bar{1},\bar{2},\bar{3}$. In addition
to $A$, we also need to know the matrix $S$ which gives the inversion
symmetry (and in general the effect of the $\mathbb{Z}_{4}$ symmetry).
In this case it exchanges the left and right copies: 
\begin{equation}
p_{\ell}(\tfrac{1}{x})=S_{\ell m}p_{m}(x),\qquad S=1_{3\times3}\otimes\left[\begin{array}{cc}
0 & 1\\
1 & 0
\end{array}\right].\label{eq:inv-cond-p}
\end{equation}
The vacuum algebraic curve has poles at $x=\pm1$, controlled by a
vector $\kappa_{\ell}$: 
\begin{equation}
p_{\ell}=\frac{\kappa_{\ell}x}{x^{2}-1},\qquad\kappa=\frac{\Delta}{2g}(0,-1,0,\:0,1,0)\,.\label{eq:vacuum-p}
\end{equation}
This must satisfy $S_{\ell m}\kappa_{m}=-\kappa_{\ell}$ and $\kappa_{\ell}A_{\ell m}\kappa_{m}=0$;
the particular solution is chosen by explicitly calculating the monodromy
matrix \cite{Zarembo:2010yz} for the BMN point particle \cite{Berenstein:2002jq}. 

Solutions above this vacuum are constructed by introducing various
cuts. The crucial equation here is that when crossing a cut $C$ (of
mode number $n$) in sheet $\ell$, the change in $p_{\ell}$ is given
by 
\begin{equation}
p_{\ell}\to p_{\ell}-A_{\ell m}p_{m}+2\pi n\,.\label{eq:jump-condition-p}
\end{equation}
When $A_{\ell\ell}=2$ this gives the change expected for a square
root cut, but this is the more general form. As we approach the branch
point, the change must go to zero, since continuity demands that it
must agree with the result of walking around the end of the cut. This
gives an equation for the positions of branch points: $2\pi n=A_{\ell m}p_{m}(x)$. 

It will be useful to also write another set of quasimomenta $q_{i}$,
corresponding to the basis directions in the following representation
of the weight vectors:
\newcommand{\mydynkin}{
\rlap{\hspace{-3mm}
\begin{tikzpicture}[scale=0.62, darkred]
\draw (0,0) -- (4,0);
\draw [fill=white] (0,0) circle (2mm); 
\draw [fill=white] (2,0) circle (2mm); 
\draw [fill=white] (4,0) circle (2mm); 
\draw (18mm,0) -- (22mm,0);
\draw (2,-2mm) -- (2,2mm);
\end{tikzpicture}
}}

\newcommand{\whitebox}{\rlap{\smash{\hspace{-15mm}\raisebox{0mm}{
\tikz \draw [white, fill=white] (0,0) rectangle (48mm,0.5);
}}}} 
\begin{equation}
\begin{array}{cccccccc|ccc}
\Lambda_{1} & \Lambda_{2} & \Lambda_{3} &  & \Lambda_{\bar{1}} & \Lambda_{\bar{2}}\; & \Lambda_{\bar{3}}\;\, &  &  & \vspace{-1mm}\\
\mydynkin & \vphantom{\dfrac{1}{1}} &  &  & \;\mydynkin &  &  &  &  & \vspace{-1mm} & i\;\\
 & -1 &  &  &  &  &  &  & F &  & 1\\
\hline 2\sin^{2}\phi & -1 & 2\cos^{2}\phi &  &  &  &  &  & B &  & 3\\
-\sin2\phi &  & \sin2\phi &  &  &  &  &  & B & \vspace{2mm} & 5\\
 &  &  &  &  & 1 &  &  & F &  & 2\\
\hline \whitebox &  &  &  & -2\sin^{2}\phi & 1 & -2\cos^{2}\phi &  & B &  & 4\\
 &  &  &  & \sin2\phi &  & -\sin2\phi &  & B &  & 6
\end{array}\label{eq:B-basis}
\end{equation}
This is a solution to $\Lambda_{\ell}\cdot\Lambda_{m}=A_{\ell m}$
which reduces to the vectors \cite{Zarembo:2010yz} gave for $\alpha=\tfrac{1}{2}$
at least on the left (i.e for the unbarred $\ell=1,2,3$). I have
inserted a minus into the right half (barred $\ell$) for later convenience.
In terms of the $q_{i}(x)$, the algebraic curve with the vacuum plus
resolvents $G_{1}$ and $G_{3}$ turned on is: 
\begin{equation}
\left(\begin{array}{c}
q_{1}\\
q_{2}\\
\hline q_{3}\\
q_{4}\\
q_{5}\\
q_{6}
\end{array}\right)=\left(\begin{array}{l}
\frac{\Delta}{2g}\frac{x}{x^{2}-1}\\
\frac{\Delta}{2g}\frac{x}{x^{2}-1}\\
\hline \frac{\Delta}{2g}\frac{x}{x^{2}-1}+2\sin^{2}\phi\, G_{1}(x)+2\cos^{2}\phi\, G_{3}(x)\\
\frac{\Delta}{2g}\frac{x}{x^{2}-1}-2\sin^{2}\phi\, G_{1}(\tfrac{1}{x})-2\cos^{2}\phi\, G_{3}(\tfrac{1}{x})\\
\qquad\quad-\sin2\phi\, G_{1}(x)+\sin2\phi\, G_{3}(x)\\
\qquad\quad+\sin2\phi\, G_{1}(\tfrac{1}{x})-\sin2\phi\, G_{3}(\tfrac{1}{x})
\end{array}\right)\to\smash{\frac{1}{2gx}\left(\begin{array}{l}
\Delta+S\\
\Delta-S\\
\hline J'-Q'\\
J'+Q'\\
Q_{5}\\
Q_{6}
\end{array}\right)+\mathcal{O}\Big(\frac{1}{x^{2}}\Big).}\label{eq:quasi-q}
\end{equation}
This is reminiscent of $AdS_{4}\times CP^{3}$ in that the bosonic
resolvents $G_{1}$ and $G_{3}$ each appear on two sheets, one of
them lacking the pole with $\Delta/g$ from the vacuum: we would call
these light modes. $G_{\bar{1}}$ and $G_{\bar{3}}$ are similar.
The terms $G(\tfrac{1}{x})$ have been filled in by the inversion
symmetry, which now reads 
\[
q_{2}(x)=-q_{1}(\tfrac{1}{x}),\qquad q_{4}(x)=-q_{3}(\tfrac{1}{x}),\qquad q_{6}(x)=-q_{5}(\tfrac{1}{x}).
\]

The global charges of the string are given by the large-$x$ behaviour
of the quasimomenta. In general let us define $J_{\ell}$ corresponding
to the Cartan generators, and $Q_{i}=J_{\ell}\Lambda_{\ell i}$: 
\[
p_{\ell}(x)\to\frac{1}{2gx}J_{\ell},\qquad q_{i}(x)\to\frac{1}{2gx}Q_{i}+\mathcal{O}\Big(\frac{1}{x^{2}}\Big)\qquad\mbox{as }x\to\infty.
\]
The right hand side of \eqref{eq:quasi-q} defines charges $\Delta,S$
from the AdS directions, and $J',Q'$ from the spheres. In terms of
$J_{\ell}$ these are \vspace{-2mm} 
\begin{align}
\Delta & =\tfrac{1}{2}(-J_{2}+J_{\bar{2}})\nonumber \\
S & =\tfrac{1}{2}(-J_{2}-J_{\bar{2}})\displaybreak[0]\nonumber \\
J' & =\sin^{2}\phi\,(J_{1}-J_{\bar{1}})-\tfrac{1}{2}(J_{2}-J_{\bar{2}})+\cos^{2}\phi\,(J_{3}-J_{\bar{3}})\nonumber \\
Q' & =-\sin^{2}\phi\,(J_{1}+J_{\bar{1}})+\tfrac{1}{2}(J_{2}+J_{\bar{2}})-\cos^{2}\phi\,(J_{3}+J_{\bar{3}})\displaybreak[0]\label{eq:def-Qprime}\\
\shortintertext{and\ we\ will\ also\ want}J & =\; J_{1}-J_{\bar{1}}\;-J_{2}+J_{\bar{2}}\;+J_{3}-J_{\bar{3}}\nonumber \\
Q & =-J_{1}-J_{\bar{1}}\;+J_{2}+J_{\bar{2}}\;-J_{3}-J_{\bar{3}}\,.\label{eq:def-Q}
\end{align}

For solutions with nonzero worldsheet momentum (i.e. solutions which
are not by themselves closed strings) we must allow the quasimomentum
to have a constant term at infinity:
\[
p_{\ell}(x)\to P_{\ell}+\frac{1}{2gx}J_{\ell}+\mathcal{O}\Big(\frac{1}{x^{2}}\Big).
\]
The total momentum is given by
\begin{equation}
P=2\sin^{2}\phi\left(-P_{1}+P_{\bar{1}}\right)+2\cos^{2}\phi\left(-P_{3}+P_{\bar{3}}\right).\label{eq:def-P}
\end{equation}

\newcommand{\smalldynkin}[3]{
\begin{tikzpicture}[scale=0.6]
\draw (1,0) -- (3,0);
\draw [fill=#1] (1,0) circle (2mm); 
\draw [fill=#2] (2,0) circle (2mm); 
\draw [fill=#3] (3,0) circle (2mm); 
\draw [thin] (18mm,0) -- (22mm,0);
\draw [thin] (2,-2mm) -- (2,2mm);
\end{tikzpicture}
} 

\newcommand{\colourone}{midgreen} \newcommand{\colourtwo}{darkred} 

\newcommand{\smallcirc}[1]{\begin{tikzpicture}[scale=0.6] \draw [fill=#1] (0,0) circle (2mm); \end{tikzpicture}}

\subsection{Constructing Modes}

The first fluctuation of the vacuum solution is given by turning on
a new pole with canonical residue $-\alpha(y)$ \cite{Beisert:2005bv}:
\[
G_{1}(x)=-\frac{\alpha(y)}{x-y}+\frac{1}{2}\frac{\alpha(y)}{-y},\qquad\alpha(y)=\frac{1}{2g}\frac{y^{2}}{y^{2}-1}.
\]
The perturbation may also alter the residues at $\pm1$, and at infinity
it must behave as follows: 
\[
\left(\begin{array}{c}
\delta q_{1}\\
\delta q_{2}\\
\hline \delta q_{3}\vphantom{\frac{\alpha(y)}{x-y}}\\
\delta q_{4}\vphantom{\frac{\alpha(y)}{x-y}}\\
\delta q_{5}\vphantom{\frac{\alpha(y)}{x-y}}\\
\delta q_{6}
\end{array}\right)=\left(\begin{array}{l}
\delta K\\
\delta K\\
\hline \delta K+2\sin^{2}\phi\:[\frac{\alpha(y)}{x-y}+\frac{\alpha(y)}{2y}]\\
\delta K-2\sin^{2}\phi\:[\frac{\alpha(y)}{1/x-y}+\frac{\alpha(y)}{2y}]\\
\delta K_{5}-\sin2\phi\:[\frac{\alpha(y)}{x-y}+\frac{\alpha(y)}{2y}]\\
\delta K_{6}+\sin2\phi\:[\frac{\alpha(y)}{1/x-y}+\frac{\alpha(y)}{2y}]
\end{array}\right)\to\frac{1}{2gx}\left(\begin{array}{l}
\delta\Delta\\
\delta\Delta\\
\hline 2\sin^{2}\phi\vphantom{\frac{\alpha(y)}{x-y}}\\
0\vphantom{\frac{\alpha(y)}{x-y}}\\
-\sin2\phi\vphantom{\frac{\alpha(y)}{x-y}}\\
0
\end{array}\right)+\mathcal{O}\Big(\frac{1}{x^{2}}\Big).
\]
Here $\delta K=\smash{\frac{\delta\Delta}{2g}\frac{x}{x^{2}-1}}$
for the first four sheets, synchronised as in \eqref{eq:vacuum-p},
and $\delta K_{5}=\delta K_{6}=b\smash{\frac{x}{x^{2}-1}}$.%
\footnote{The pole $\delta K_{5}=\delta K_{6}$ corresponds to $\kappa'=\tfrac{1}{2}b(-\cot\phi,0,\tan\phi,\:\cot\phi,0,-\tan\phi)$
which like \eqref{eq:vacuum-p} is a $-1$ eigenvector of $S$. The
solution has $b=\frac{\sin2\phi}{2g}\frac{1}{1-y^{2}}$.%
} From the sheets connected by the new poles in $\delta q(x)$ we might
call this the $(3,5)$ mode. Solving the conditions at infinity, we
get off-shell frequency 
\begin{equation}
\delta\Delta=\Omega_{1}(y)=\frac{2\sin^{2}\phi}{y^{2}-1}.\label{eq:1mode-Ohm}
\end{equation}
The perturbation carries some momentum,%
\footnote{This momentum is often avoided by considering a pair of fluctuations
at $\pm y$ as in \cite{Gromov:2008bz,Bandres:2009kw}, see also \cite{Abbott:2010yb}.
Doing so for the giant magnon, you would miss the second term in \eqref{eq:magnon-Ohm}.%
} and solving the inversion conditions gives $P_{1}=-P_{\bar{1}}\neq0$
and
\begin{equation}
\delta P=\frac{\sin^{2}\phi}{g}\frac{y}{y^{2}-1}.\label{eq:1mode-p}
\end{equation}

Now to find the position of this mode in the spectral plane, we take
\eqref{eq:jump-condition-p} and demand continuity as we approach
the end of the infinitesimally short branch cut which we are inserting.
This gives%
\footnote{Compared to $q_{3}-q_{5}=(2\sin^{2}\phi+\sin2\phi)p_{1}-p_{2}+(2\cos^{2}\phi-\sin2\phi)p_{3}$
we see that \eqref{eq:1mode-twopin} reduces to $2\pi n=q_{3}-q_{5}$
as in \cite{Gromov:2007aq} only at $\phi=\frac{\pi}{4}$. The same
is true for the other modes. %
} 
\begin{align}
2\pi n_{1} & =A_{1k}p_{k}=4\sin^{2}\phi\: p_{1}-2\sin^{2}\phi\: p_{2}\label{eq:1mode-twopin}\\
 & =2\sin^{2}\phi\:\frac{\Delta}{2g}\frac{x}{x^{2}-1}\nonumber 
\end{align}
using the vacuum solution \eqref{eq:vacuum-p} on the second line.
Solving for $x$, and choosing the solution outside the unit circle,
we get 
\[
x_{n}=\frac{\Delta\sin^{2}\phi}{4\pi\, g\, n}\pm\sqrt{1+\Big(\frac{\Delta\sin^{2}\phi}{4\pi\, g\, n}\Big)^{2}}
\]
giving the desired on-shell frequency 
\begin{equation}
\omega_{n}=\Omega_{1}(x_{n})=-\sin^{2}\phi+\sqrt{\sin^{4}\phi+\Big(\frac{4\pi\, g\, n}{\Delta}\Big)^{2}}.\label{eq:1mode-w_n}
\end{equation}
The charges \eqref{eq:def-Qprime}, \eqref{eq:def-Q} are
\[
\delta J=-1,\qquad\delta Q=+1,\qquad\delta J'=-\sin^{2}\phi,\qquad\delta Q'=\sin^{2}\phi
\]
and the momentum \eqref{eq:1mode-p} is $\delta P=2\pi n/\Delta$,
so that if $\omega=\Delta-J'-Q'$ this matches \eqref{eq:one-magnon-disp}. 

The construction of all the other modes is similar:
\begin{enumerate}
\item [1$f$.] \setcounter{enumi}{2}A fermion of the same mass is obtained
by turning on $G_{1}(x)=G_{2}(x)=\smash{-\frac{\alpha(y)}{x-y}+\frac{1}{2}\frac{\alpha(y)}{-y}}$.
 The equation for the perturbation can be written 
\[
\left(\begin{array}{c}
\delta q_{1}\\
\hline \delta q_{3}\\
\delta q_{5}
\end{array}\right)=\left(\begin{array}{l}
\delta K+[\frac{\alpha(y)}{x-y}+\frac{\alpha(y)}{2y}]\\
\hline \delta K+(1-2\sin^{2}\phi)\:[\frac{\alpha(y)}{x-y}+\frac{\alpha(y)}{2y}]\\
\delta K_{5}+\sin2\phi\:[\frac{\alpha(y)}{x-y}+\frac{\alpha(y)}{2y}]
\end{array}\right)\to\frac{1}{2gx}\left(\begin{array}{l}
\delta\Delta+1\\
\hline 1-2\sin^{2}\phi\\
\sin2\phi
\end{array}\right)+\ldots
\]
(with the rest filled in by inversion symmetry as before). At $\phi=\frac{\pi}{4}$
this has new poles on just two sheets and thus might be called the
$(1,5)$ mode, but for general $\phi$ this interpretation is not
clear; let us call it ``$1f$''. The resulting off-shell frequency
is the same, $\Omega_{1f}(y)=\Omega_{1}(y)$. The positions of the
poles come from 
\begin{align}
2\pi n_{1f} & =\sum_{\ell=1,2}A_{\ell k}p_{k}\label{eq:1fmode-2pin}\\
 & =2\sin^{2}\phi\: p_{1}-2\sin^{2}\phi\: p_{2}-2\cos^{2}\phi\: p_{3}\nonumber \\
 & =2\sin^{2}\phi\:\frac{\Delta}{2g}\frac{x}{x^{2}-1}\qquad\mbox{(for the vacuum).}\nonumber 
\end{align}
Thus we will get exactly the same frequencies $\omega_{n}$ as for
the boson. 
\begin{table}
\centering %
\begin{tabular}{cccccc|c}
$r$ & $m_{r}$ &  & $2\pi n_{r}$ & $\delta J$ & $\delta Q$ & $\alpha=\tfrac{1}{2}$ i.e. $\phi=\tfrac{\pi}{4}$\tabularnewline
\hline 
$1$, $\bar{1}$ & $\sin^{2}\phi$ & \smalldynkin{\colourone}{white}{white} & $A_{1k}p_{k}$ , or $-A_{\bar{1}k}p_{k}$ & $-1$ & $1$, $-1$ & $(3,5)$, $(4,6)$\tabularnewline
$1f$, $\bar{1}f$ & $\sin^{2}\phi$ & \smalldynkin{\colourone}{\colourone}{white} & $(A_{1k}+A_{2k})p_{k}$ , or sim.  & 0 & 0 & $(1,5)$, $(2,6)$\tabularnewline
$3$, $\bar{3}$ & $\cos^{2}\phi$ & \smalldynkin{white}{white}{\colourone} & $A_{3k}p_{k}$ & $-1$ & $1$, $-1$ & $(3,-5)$, $(4,-6)$\tabularnewline
$3f$, $\bar{3}f$ & $\cos^{2}\phi$ & \smalldynkin{white}{\colourone}{\colourone} & $(A_{2k}+A_{3k})p_{k}$ & 0 & 0 & $(1,-5)$, $(2,-6)$\tabularnewline
\hline 
$4$, $\bar{4}$ & 1 & \smalldynkin{\colourone}{\colourtwo}{\colourone} & $(A_{1k}+2A_{2k}+A_{3k})p_{k}$ & 0 & 0 & $(1,-1)$, $(2,-2)$\tabularnewline
$4f$, $\bar{4}f$ & 1 & \smalldynkin{\colourone}{\colourone}{\colourone} & $(A_{1k}+A_{2k}+A_{3k})p_{k}$ & $-1$ & $1$, $-1$ & $(1,-3)$, $(2,-4)$\tabularnewline
\hline 
\end{tabular}\caption[Fake caption without tikz figures.]{List of modes in the $AdS_{3}\times S^{3}\times S^{3}$ algebraic
curve. The colouring of the nodes is $-k_{\ell r}$ with $\smallcirc{\colourone}= +1, -1$
and $\smallcirc{\colourtwo}= +2, -2$, writing ``left, right'' everywhere.\label{tab:List-of-modes} }
\end{table}

\item We can treat the boson ``$3$'' (from $G_{3}=\smash{-\frac{\alpha(y)}{x-y}+\frac{1}{2}\frac{\alpha(y)}{-y}}$)
and the fermion ``$3f$'' in exactly the same way, obtaining\vspace{-2mm} 
\begin{align*}
\Omega_{3}(y) & =\Omega_{3f}(y)=\frac{2\cos^{2}\phi}{y^{2}-1}\displaybreak[0]\\
\omega_{n} & =-\cos^{2}\phi+\sqrt{\cos^{4}\phi+\Big(\frac{4\pi\, g\, n}{\Delta}\Big)^{2}}.
\end{align*}
Note that $\delta q_{5}$ has the opposite sign for these modes compared
to $1,1f$ above; at $\phi=\tfrac{\pi}{4}$ we might therefore call
them $(3,-5)$ and $(1,-5)$, thinking of $q_{-i}=-q_{i}$ as another
six sheets. 
\item The heavy modes can be constructed by simply adding two light modes:
$4=1f+3f$, and $4f=1+3f=3+1f$. This addition is at the level of
$\delta q_{i}(x)$, and hence applies to $\Omega_{r}(y)$ too: 
\[
\Omega_{4}(y)=\Omega_{4f}(y)=\frac{2}{y^{2}-1}.
\]
For the mode numbers, we have%
\footnote{As in $AdS_{4}\times CP^{3}$ the heavy modes correspond to stacks
of Bethe roots \cite{Beisert:2005di,Gromov:2007ky}. In general few
solutions $x$ here will make $n_{1f}$ and $n_{3f}$ integers, but
in the thermodynamic limit $\Delta/g\to\infty$ this constraint disappears. %
} 
\begin{align}
2\pi n_{4} & =(A_{1k}+2A_{2k}+A_{3k})p_{k}=2\pi n_{1f}+2\pi n_{3f}\label{eq:4mode-2pin}\\
 & =-2p_{2}=2\frac{\Delta}{2g}\frac{x}{x^{2}-1}\nonumber 
\end{align}
and thus 
\[
\omega_{n}=-1+\sqrt{1+\Big(\frac{4\pi\, g\, n}{\Delta}\Big)^{2}}.
\]
For the heavy boson we must add two fermions, not two bosons: $4\neq1+3$.
This and $\bar{4}$ are the two transverse direction in $AdS_{3}$;
unlike the $CP^{3}$ case there are no heavy bosons in the sphere
directions. 
\item [$\bar{1}$.] Finally, the barred modes differ only by some minus
signs: $G_{\bar{1}}(x)=\smash{+\frac{\alpha(y)}{x-y}-\frac{1}{2}\frac{\alpha(y)}{-y}}$
and $2\pi n_{\bar{1}}=-A_{\bar{1}k}p_{k}$, and $\delta Q=-1$. But
\eqref{eq:1mode-Ohm} and \eqref{eq:1mode-w_n} are the same. \medskip\\
The addition of modes used above is one half of \cite{Gromov:2008ec}'s
efficient procedure for constructing off-shell frequencies, and these
barred modes may be constructed by the other half, namely the use
of the inversion symmetry. Writing the perturbation for mode $r$
(with a pole at $y$)  as $\smash{\delta_{r}^{(y)}}p_{\ell}(x)$,
the new pole for the corresponding mode on the right is given by $\smash{\delta_{\overline{r}}^{(y)}}p_{\ell}(x)=-\smash{\delta_{r}^{(1/y)}}p_{\ell}(x)$,
exactly as in \cite{Gromov:2008ec}. 
\end{enumerate}
Table \ref{tab:List-of-modes} summarises some properties of these
modes. Another summary is as follows: The perturbation $\delta q_{i}$
for $N_{r}$ excitations of the mode $r$ has a pole at $y$ with
residue $k_{ir}N_{r}\alpha(y)$, where $k_{ir}$ are the coefficients
inside the bracket below. This condition at infinity constrains the
perturbation sufficiently that we can read off $\Omega(y)=\delta\Delta$:
\begin{equation}
\left(\begin{array}{c}
\delta q_{1}\\
\delta q_{2}\\
\hline \delta q_{3}\\
\delta q_{4}\\
\delta q_{5}\\
\delta q_{6}
\end{array}\right)\to\frac{1}{2gx}\left(\begin{array}{l}
\delta\Delta+2N_{4}+N_{1f}+N_{3f}+N_{4f}\\
\delta\Delta+2N_{\bar{4}}+N_{\bar{1}f}+N_{\bar{3}f}+N_{\bar{4}f}\\
\hline -2\sin^{2}\phi\:(N_{1}+N_{1f})+N_{1f}-2\cos^{2}\phi\:(N_{3}+N_{3f})+N_{3f}-N_{4f}\\
-2\sin^{2}\phi\:(N_{\bar{1}}+N_{\bar{1}f})+N_{\bar{1}f}-2\cos^{2}\phi\:(N_{\bar{3}}+N_{\bar{3}f})+N_{\bar{3}f}-N_{\bar{4}f}\\
\sin2\phi\:(N_{1}+N_{1f}-N_{3}-N_{3f})\\
\sin2\phi\:(N_{\bar{1}}+N_{\bar{1}f}-N_{\bar{3}}-N_{\bar{3}f})
\end{array}\right)+\ldots\label{eq:dq-N-infinity}
\end{equation}
The analogous equation in terms of $p_{\ell}$ has $\delta p_{\ell}\to\frac{1}{2gx}(-\tfrac{1}{2},-1,-\tfrac{1}{2},\;\tfrac{1}{2},1,\tfrac{1}{2})\delta\Delta+\frac{1}{2gx}k_{\ell r}N_{r}$
as $x\to\infty$, with $k_{\ell r}=\pm1,\pm2$ taken from the colouring
in table \ref{tab:List-of-modes}.

\section{Summation Prescriptions\label{sec:Defining-cutoff-prescriptions}}

Semiclassical quantisation gives the one-loop correction to the energy
of a soliton as 
\begin{equation}
\delta E=\sum_{r}\sum_{n=-\infty}^{\infty}(-1)^{F_{r}}\frac{1}{2}\omega_{n}^{r}-\delta E_{\mathrm{vac}}\,.\label{eq:sum-w-unregulated}
\end{equation}
For us $\delta E_{\mathrm{vac}}=0$. The sum for any one polarisation
$r$ will diverge quadratically, but for a matched set of bosons and
fermions cancellations typically tame this to a logarithmic divergence.
This still leaves some room for dependence on how we cut off the sum
on $n$ in the UV. 

In terms of the spectral plane, a very high-energy mode is one located
very close to $x=1$, with energy 
\[
\omega=\Omega_{r}(1+\epsilon)=\frac{2m_{r}}{\epsilon(2+\epsilon)}=\frac{m_{r}}{\epsilon}+\mathcal{O}(\epsilon^{0}).
\]
In terms of mode numbers, instead 
\[
\omega_{N}=-m_{r}+\sqrt{m_{r}^{2}+\Big(\frac{4\pi gN}{\Delta}\Big)^{2}}=N\frac{4\pi g}{\Delta}-m_{r}+\mathcal{O}\Big(\frac{1}{N}\Big).
\]
The principal options for how to regulate the modes of different masses
follow from these:
\begin{enumerate}
\item [i.] A cutoff at a fixed physical energy $\omega=\Lambda$ is a cutoff
at the same mode number for all polarisations (provided there are
no divergences stronger than $\log N$), but at radius $x=1+m_{r}/\Lambda$
in the spectral plane for a mode of mass $m_{r}$. It is what would
seem most natural to a resident of the target space ignorant of integrability,
and is sometimes referred to as the worldsheet prescription, although
(as we showed in \cite{Abbott:2010yb}) can quite easily be implemented
in the algebraic curve description.%
\footnote{Observe also that this ``physical'' prescription corresponds also
to a cutoff in worldsheet momentum, to the precision required here.
At $x=1+\epsilon$ we have from \eqref{eq:1mode-p} and the equivalent
for the $m_{3}$ modes $P\propto\omega+\mathcal{O}(\epsilon^{0})$.%
}
\item [ii.] The alternative is a cutoff at fixed radius in the spectral
plane, which corresponds to a cutoff at mode number $N_{r}=m_{r}\frac{\Delta}{4\pi g}\frac{1}{\epsilon}$
. This is certainly the easiest to implement in the algebraic curve
language, although it can clearly also be used to add up frequencies
from a worldsheet calculation, as Gromov and Mikhaylov\cite{Gromov:2008fy}
did upon introducing this idea for $AdS_{4}\times CP^{3}$. To do
so however we still need to identify the heavy modes, which for classical
solutions far from the BMN vacuum is not necessarily obvious. For
this the algebraic curve is definitive: the heavy modes are those
whose off-shell perturbation is the sum of two light modes'. \medskip\\
In the present $AdS_{3}\times S^{3}\times S^{3}\times S^{1}$ case
this ``new'' prescription leads to three different cutoffs in terms
of energy, or mode number, and these clearly change as $\alpha=\cos^{2}\phi$
is changed. It will be important below that as we approach $\alpha=1$,
where one of the spheres decompactifies, the cutoff for the mode which
is becoming massless drops to zero, completely excluding this mode
from the sum. (See also the integral form \eqref{eq:dE-new-integral}
below.) 
\end{enumerate}
One argument advanced in favour of the new prescription in $AdS_{4}\times CP^{3}$
involves the fact that the total energy of a pair of light modes exactly
at their cutoffs is the same as that of the corresponding heavy mode
at its cutoff \cite{Astolfi:2011ju,Astolfi:2011bg}. This is still
true here as each heavy mode is made of two light modes of opposite
mass, and $m_{1}+m_{3}=m_{4}$. There is a third option which also
has this property:
\begin{enumerate}
\item [iii.] We could simply cut off both light modes at half the energy
of the heavy mode. In terms of mode numbers and the spectral plane
this means 
\begin{equation}
N_{1}=N_{3}=\tfrac{1}{2}N_{4},\qquad\begin{aligned}\epsilon_{1} & =2m_{1}\epsilon_{4}\\
\epsilon_{3} & =2m_{3}\epsilon_{4}\,.
\end{aligned}
\label{eq:third-way}
\end{equation}
For $\alpha=\tfrac{1}{2}$ (and for $AdS_{4}$) this is  identical
to the new prescription. It avoids turning off the newly massless
modes $m_{1}$ as we approach $\alpha=1$, but once we get there still
treats the modes $m_{3}=1$ and $m_{4}=1$ differently. For this reason
it seems undesirable, and I will mention it only as an afterthought. 
\end{enumerate}

\section{Giant Magnons\label{sec:Giant-Magnons}}

Giant magnons are the macroscopic classical string solutions corresponding
to elementary excitations with momentum $p$ of order 1 \cite{Hofman:2006xt}.
Bound states of a large number $Q\sim g$ of magnons form dyonic giant
magnons \cite{Dorey:2006dq,Chen:2006gea}; from the algebraic point
of view this is the natural case, and they are described by a single
log cut \cite{Minahan:2006bd,Vicedo:2007rp,Chen:2007vs}. The branch
points of this are the Zhukovsky variables $\Xpm$, which are defined
here by \cite{OhlssonSax:2011ms} 
\[
\Xp+\frac{1}{\Xp}-\Xm-\frac{1}{\Xm}=i\frac{2m_{r}}{h}Q,\qquad\frac{\Xp}{\Xm}=e^{ip}
\]
with $Q=1$ for an elementary excitation. The exact dispersion relation
is 
\begin{equation}
E(p)=-i\frac{h}{2}\Big(\Xp-\frac{1}{\Xp}-\mbox{c.c.}\Big)=\sqrt{Q^{2}m_{r}^{2}+4h^{2}\sin^{2}\frac{p}{2}}.\label{eq:disp-rel-Q}
\end{equation}
At strong coupling we can expand this using $h=2g+c+\mathcal{O}(1/g)$
to get the classical energy and one-loop correction $E_{0}+\delta E+\ldots$.
This is to be done holding $p$ and $Q$ fixed. 

The classical solution for a magnon on the first sphere is \eqref{eq:quasi-q}
with 
\begin{equation}
G_{1}(x)=\frac{1}{2\sin^{2}\phi}\Big[G_{\mathrm{mag}}(x)-\tfrac{1}{2}G_{\mathrm{mag}}(0)\Big],\qquad G_{\mathrm{mag}}(x)=-i\log\Big(\frac{x-\Xp}{x-\Xm}\Big).\label{eq:G1-Gmag}
\end{equation}
The prefactor is needed to cancel the non-integer factor in \eqref{eq:1mode-twopin},
from \eqref{eq:jump-condition-p}, and also in \eqref{eq:def-P}.
The asymptotic charges \eqref{eq:def-Qprime}, \eqref{eq:def-Q} for
this are
\begin{align*}
J' & =\Delta+ig\left(\Xp-\frac{1}{\Xp}-\mbox{c.c.}\right)\displaybreak[0]\\
Q' & =\sin^{2}\phi\, Q=-ig\left(\Xp+\frac{1}{\Xp}-\mbox{c.c.}\right)\displaybreak[0]\\
\shortintertext{and\ the\ momentum\ is}P & =G_{\mathrm{mag}}(0)=p.
\end{align*}
 This clearly gives precisely the desired dispersion relation with
$E_{0}=\Delta-J'$. 

There is of course a similar magnon on the second sphere, with $G_{3}=\smash{\frac{1}{2\cos^{2}\phi}}[G_{\mathrm{mag}}(x)-\tfrac{p}{2}]$,
a giant version of the mode ``$3$''. Note however that there is
no analogue of the $RP^{3}$ giant magnon in $AdS_{4}\times CP^{3}$,
in which turning on magnons in two sectors led to a simplification.
That solution was was a giant version of the $CP^{3}$ heavy boson;
here the only heavy bosons are the $AdS$ modes $4$ and $\bar{4}$.
(See however footnote \ref{fn:non-cromulent} in the conclusions.)

\subsection{One-loop Correction}

To calculate the one-loop correction to the energy we should begin
by finding the off-shell frequencies, by constructing the perturbations
of the quasimomenta. The two differences from the BMN modes above
are that here is that we must allow the endpoints of the cut to move,
and that we do not allow the perturbation to alter the total momentum.
For instance for the $3f$ mode and the magnon \eqref{eq:G1-Gmag},
the perturbation must obey 
\[
\negthickspace\delta q=\negthickspace\left(\negthickspace\begin{array}{l}
\delta K\\
\delta K\\
\hline \delta K+(1-2\cos^{2}\phi)\:[\frac{\alpha(y)}{x-y}+\frac{\alpha(y)}{2y}]+2\sin^{2}\phi\:[H(x)-\tfrac{1}{2}H(0)]\\
\delta K+(1-2\cos^{2}\phi)\:[\frac{\alpha(y)}{1/x-y}+\frac{\alpha(y)}{2y}]-2\sin^{2}\phi\:[H(\tfrac{1}{x})-\tfrac{1}{2}H(0)]\negthickspace\negthickspace\\
\delta K_{5}-\sin2\phi\:[\frac{\alpha(y)}{x-y}+\frac{\alpha(y)}{2y}]-\sin2\phi\:[H(x)-\tfrac{1}{2}H(0)]\\
\delta K_{6}-\sin2\phi\:[\frac{\alpha(y)}{1/x-y}+\frac{\alpha(y)}{2y}]+\sin2\phi\:[H(\tfrac{1}{x})-\tfrac{1}{2}H(0)]
\end{array}\right)\negthickspace\to\frac{1}{2gx}\negthickspace\left(\negthickspace\begin{array}{l}
\delta\Delta\\
\delta\Delta\\
\hline 1-2\cos^{2}\phi\vphantom{\frac{\alpha(y)}{x-y}}\negthickspace\negthickspace\\
0\vphantom{\frac{\alpha(y)}{x-y}}\\
-\sin2\phi\vphantom{\frac{\alpha(y)}{x-y}}\\
0
\end{array}\right)\negthickspace+\ldots
\]
where $H(x)=\frac{A_{+}}{x-\Xp}+\frac{A_{-}}{x-\Xm}$ has been inserted
wherever the classical solution has $G_{\mathrm{mag}}(x)$. The resulting
frequency is the $r=3f$ case of 
\begin{equation}
\Omega_{r}(y)=\frac{2\, m_{r}}{y^{2}-1}\left(1-y\frac{\Xp+\Xm}{\Xp\Xm+1}\right)=m_{r}\,\Omega_{\mathrm{mag}}(y).\label{eq:magnon-Ohm}
\end{equation}
The same formula applies to all light and heavy modes, and equally
for the magnon in $G_{3}$. In fact it is the same as in $S^{5}$
and $CP^{3}$, and as discussed in \cite{Gromov:2008ie} the first
term can be interpreted as the energy of the mode itself \eqref{eq:1mode-Ohm},
and the second as the effect of adjusting the magnon's momentum to
compensate for the momentum \eqref{eq:1mode-p} carried by the perturbation.

While we cannot write the frequencies $\omega_{n}$ in closed form,
we can still compute the one-loop correction. This is done by using
\cite{SchaferNameki:2006gk}'s trick to write it as a contour integral
in $n$: 
\[
\delta E=\sum_{n,r}\frac{1}{2}\omega_{n}^{r}=\frac{1}{4i}\oint dn\,\sum_{r}(-1)^{F_{r}}\cot(\pi n)\:\Omega_{r}(x_{n}^{r}).
\]
For each $r$ we can now use the relevant equation for the position
of the pole to write this as an integral in $x$, along a contour
$W$ enclosing poles along the real line at $\left|x\right|>1$. Then
we deform this contour to the unit circle, $-U$ counting orientation:
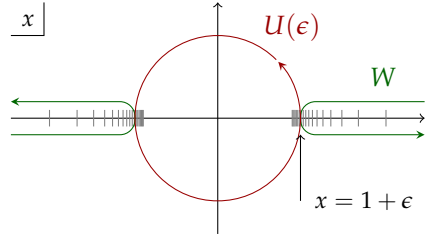
\[
\begin{tikzpicture}[scale=1.1] 
\draw [->] (-2.5,0) -- (2.5,0); 
\draw [->] (0,-1.4) -- (0,1.4); 
\draw (-2.5,1) -- (-2.1,1) -- (-2.1,1.4); 
\node [anchor=south east] at (-2.1,1) {\small $x$};

\foreach \x in {3,...,20} 
  {
  \draw [gray] (0.7+4/\x,-1mm) -- (0.7+4/\x,1mm);  
  \draw [gray] (-0.7-4/\x,-1mm) -- (-0.7-4/\x,1mm);
  };

\draw [stealth-, rounded corners=2mm, darkgreen] (2.5,-0.2) -- (1,-0.2) -- (1,0.2) -- (2.5,0.2);
\draw [-stealth, rounded corners=2mm, darkgreen] (-2.5,-0.2) -- (-1,-0.2) -- (-1,0.2) -- (-2.5,0.2);
\node [darkgreen] at (2,0.5) {$W$};

\draw [-stealth, darkred] (0.707107,0.707107) arc (45:360:1cm) arc (0:43:1cm);
\node [darkred] at (0.9,1.1) {$U(\epsilon)$};

\draw [<-] (1,-2mm) -- (1,-1) node [anchor=west] {\small\negthickspace\negthickspace$x=1+\epsilon$};

\end{tikzpicture}
\vspace{-2mm}
\]  Here we can approximate $\cot(\pi n)\approx\cot(m_{r}\frac{\Delta}{g}\frac{x}{x^{2}-1})\approx\pm i$
on the upper/lower semicircles $U_{\pm}$, whose contributions are
equal, to write: 
\begin{align}
\delta E & =\frac{1}{4i}\oint_{W}dx\sum_{r}(-1)^{F_{r}}\cot(\pi n_{r})\:\partial_{x}n_{r}(x)\:\Omega_{r}(x)\label{eq:dE-with-cot}\\
 & \approx\frac{-1}{4\pi}\int_{U_{+}}\sum_{r}(-1)^{F_{r}}2\pi\partial_{x}n_{r}(x)\:\Omega_{r}(x).\nonumber 
\end{align}
 Since $\Omega_{r}=m_{r}\Omega_{\mathrm{mag}}$, let us deal with
$n_{r}$ by considering the sum over polarisations one mass at a time.
For $m_{r}=\sin^{2}\phi$ we see 
\begin{align*}
\smash{\sum_{r=1,1f,\bar{1},\bar{1}f}}(-1)^{F_{r}}2\pi n_{r} & =2\sin^{2}\phi\: p_{1}+2\cos^{2}\phi\: p_{3}-2\sin^{2}\phi\: p_{\bar{1}}-2\cos^{2}\phi\: p_{\bar{3}}\\
 & =G_{\mathrm{mag}}(x)-G_{\mathrm{mag}}(\tfrac{1}{x}).
\end{align*}
For the other masses, from table \ref{tab:List-of-modes}  it is
clear that this sum for the $m_{r}=\cos^{2}\phi$ modes is identical,
while that for $m_{r}=1$ is exactly minus this. Finally we should
allow a different cutoff for each mass, giving us the same log-divergent
integral thrice: 
\begin{align*}
\delta E & =\frac{-1}{4\pi}\Bigg\{\sin^{2}\phi\int_{U_{+}(\epsilon_{1})}\negthickspace\negthickspace dx+\cos^{2}\phi\int_{U_{+}(\epsilon_{3})}\negthickspace\negthickspace dx-\int_{U_{+}(\epsilon_{4})}\negthickspace\negthickspace dx\Bigg\}\:\partial_{x}\big[G_{\mathrm{mag}}(x)-G_{\mathrm{mag}}(\tfrac{1}{x})\big]\:\Omega_{\mathrm{mag}}(x)\\
 & =\frac{i}{\pi}\frac{\Xp-\Xm}{\Xp\Xm+1}\left[\sin^{2}\phi\:\log\epsilon_{1}+\cos^{2}\phi\:\log\epsilon_{3}-\log\epsilon_{4}\right].
\end{align*}
Clearly this will vanish for the new prescription, with all $\epsilon_{r}$
the same. For the physical prescription $\epsilon_{r}=m_{r}/\Lambda$
and (taking the non-dyonic limit $\Xpm=e^{\pm ip/2}$) instead 
\[
\delta E=\frac{1}{\pi}\sin\frac{p}{2}\;\left[\sin^{2}\phi\:\log(\sin^{2}\phi)+\cos^{2}\phi\:\log(\cos^{2}\phi)\:\right].
\]
Since $\delta E=2\, c\:\sin\frac{p}{2}$ this implies $c_{\mathrm{new}}=0$
and 
\begin{align}
c_{\mathrm{phys}} & =\frac{\sin^{2}\phi\:\log(\sin^{2}\phi)+\cos^{2}\phi\:\log(\cos^{2}\phi)}{2\pi}\label{eq:c-magnon}\\
 & =\begin{cases}
-\frac{\log2}{2\pi} & \phi=\frac{\pi}{4}\\
0 & \phi=0\,.
\end{cases}\nonumber 
\end{align}
Using the third prescription, \eqref{eq:third-way}, gives $c_{\mathrm{third}}=c_{\mathrm{phys}}+\frac{\log2}{2\pi}$;
as expected this matches $c_{\mathrm{new}}$ at $\phi=\tfrac{\pi}{4}$. 

For the dyonic case the calculation works equally well, provided we
expand the dispersion relation holding fixed not $\Xpm$ but rather
$p$ and $Q$ \cite{Abbott:2010yb}: 
\begin{equation}
\delta E=c\frac{\partial E_{0}}{\partial h}\Big\vert_{Q,p}=-2i\frac{\Xp-\Xm}{\Xp\Xm+1}c\,.\label{eq:dE-dyonic}
\end{equation}

\section{Long Spinning Strings\label{sec:Long-Spinning-Strings}}

The aim of this section is to do what Gromov and Mikhaylov \cite{Gromov:2008fy}
did for $AdS_{4}\times CP^{3}$, that is, to classify the modes of
long spinning strings as heavy or light based on the algebraic curve
description, and then to sum their frequencies using the ``new''
prescription suggested by this formalism. 

As reviewed in section \ref{sec:Defining-cutoff-prescriptions}, the
new prescription cuts off at the same radius $1+\epsilon$ in the
spectral plane for all modes, which corresponds mode number $N_{r}=m_{r}N$
for modes of mass $m_{r}$. Given the explicit frequencies $\omega_{n}$
(and since $\kappa\gg1$) we can write the sum as the following integral:\vspace{-3mm}
\begin{equation}
\delta\Delta_{\mathrm{new}}=\sum_{r}(-1)^{F_{r}}\int_{-m_{r}N}^{m_{r}N}dn\:\frac{1}{2}\omega_{n}^{r}=\int_{-N}^{N}dn\:\sum_{r}(-1)^{F_{r}}\frac{1}{2}m_{r}\omega_{m_{r}n}^{r}\,.\label{eq:dE-new-integral}
\end{equation}

The classical solution of interest is a folded string spinning in
$AdS_{3}$, stretched all the way to the boundary, times a point particle
with some momentum along the equators of both spheres. (See \eqref{eq:WS-spinning-string}
in appendix \ref{sec:Worldsheet-Giant-Magnons}.) For this solution
Forini, Puletti and Ohlsson Sax \cite{Forini:2012bb} calculated the
mode frequencies from the Green--Schwartz action; these are listed
in table \ref{tab:Modes-from-VVO}. Some comments:
\begin{itemize}
\item The parameter $\nu=J'/4\pi g$ controls the momentum on the spheres,
and we are interested in two limits. First, setting $\nu=\kappa$
takes us to the BMN limit, and it is here that the mass $m_{r}$ relevant
for classifying the modes is literally the mass of the excitation,
as in \eqref{eq:1mode-w_n}. Second, sending $\nu\to0$ gives us the
simplest rotating string entirely in $AdS_{3}$, whose frequencies
should be independent of $\alpha$. 
\item In the first two columns only $\omega_{5,6}^{F}$ and $\omega_{7,8}^{F}$
assume $\alpha=\frac{1}{2}$. We know that the light fermions $5,6$
cannot both be $r=1f$ (or both $r=3f$) as this would break the symmetry
between the two spheres; this explains the classification in the last
column. 
\end{itemize}
\begin{table}
\begin{small}
\newcommand{\blabel}{ \llap{\makebox[1.6cm][l]{Bosons:}} }
\newcommand{\flabel}{ \llap{\makebox[1.6cm][l]{Fermions:}} }
\begin{align*}
 &  &  & \mbox{``BMN'' }\nu\to\kappa: &  & \mbox{``AdS'' }\nu\to0: &  & r,\:\bar{r}\\
\hline \vphantom{1^{1}}\blabel\\
\omega_{1,2}^{B} & =\left|n\right| &  & \to\left|n\right| &  & \to\left|n\right| &  & \times\\
\omega_{3,4}^{B} & =\sqrt{n^{2}+\alpha^{2}\nu^{2}}\vphantom{\frac{1}{1}} &  & \to\sqrt{n^{2}+\alpha^{2}\kappa^{2}} &  & \to\left|n\right| &  & 3\\
\omega_{5,6}^{B} & =\sqrt{n^{2}+(1-\alpha)^{2}\nu^{2}} &  & \to\sqrt{n^{2}+(1-\alpha)^{2}\kappa^{2}} &  & \to\left|n\right| &  & 1\\
\omega_{7,8}^{B} & =\sqrt{n^{2}+2\kappa^{2}\mp2\sqrt{n^{2}\nu^{2}+\kappa^{4}}} &  & \to\pm\kappa^{2}+\sqrt{n^{2}+\kappa^{2}} &  & \to\begin{cases}
\left|n\right|\\
\sqrt{n^{2}+4\kappa^{2}}
\end{cases}\displaybreak[0] &  & 4\\
\flabel\\
\omega_{1,2}^{F} & =\pm\frac{\nu}{2}+\left|n\right| &  & \to\pm\tfrac{\kappa}{2}+\left|n\right| &  & \to\left|n\right| &  & \times\\
\omega_{5,6}^{F}\big\vert_{\alpha=1/2} & =\sqrt{n^{2}+\tfrac{\kappa^{2}}{2}+\sqrt{n^{2}\nu^{2}+\tfrac{\kappa^{4}}{4}}} &  & \to\tfrac{\kappa}{2}+\sqrt{n^{2}+\tfrac{\kappa^{2}}{4}} &  & \to\sqrt{n^{2}+\kappa^{2}} &  & 1f,3f\;\\
\omega_{7,8}^{F}\big\vert_{\alpha=1/2} & =\sqrt{n^{2}+\tfrac{\kappa^{2}}{2}-\sqrt{n^{2}\nu^{2}+\tfrac{\kappa^{4}}{4}}} &  & \to-\tfrac{\kappa}{2}+\sqrt{n^{2}+\tfrac{\kappa^{2}}{4}} &  & \to\left|n\right| &  & 1f,3f\\
\vphantom{\frac{1}{1_{1_{1}}}}\omega_{3,4}^{F} & =\pm\frac{\nu}{2}+\sqrt{n^{2}+\kappa^{2}} &  & \to\pm\tfrac{\kappa}{2}+\sqrt{n^{2}+\kappa^{2}} &  & \to\sqrt{n^{2}+\kappa^{2}} &  & 4f\\
\hline 
\end{align*}
\end{small}\vspace{-1cm}

\caption{Modes of the folded spinning string in $\adssss$, taken from \cite{Forini:2012bb}'s
appendix C. \label{tab:Modes-from-VVO}}
\end{table}
It is now very simple to put these modes into an integral like that
above. Note that while the frequencies here (at $\nu=0$) are independent
of $\phi$, the cutoffs are not. Allowing at first three different
cutoffs, the result is: 
\begin{align}
\delta\Delta & =\sum_{m_{r}=\sin^{2}\phi}\negthickspace\negthickspace(-1)^{F_{r}}\sum^{N_{1}}\tfrac{1}{2}\omega_{n}^{r}+\negthickspace\sum_{m_{r}=\cos^{2}\phi}\negthickspace\negthickspace(-1)^{F_{r}}\sum^{N_{3}}\tfrac{1}{2}\omega_{n}^{r}+\sum_{m_{r}=1}(-1)^{F_{r}}\sum^{N_{4}}\tfrac{1}{2}\omega_{n}^{r}\negthickspace\negthickspace\negthickspace\displaybreak[0]\nonumber \\
 & =\kappa\left[-\log4+\log N_{4}-\tfrac{1}{2}\log N_{3}-\tfrac{1}{2}\log N_{1}\right]\nonumber \\
 & =\begin{cases}
-\kappa\log4 & \mbox{using the physical prescription}\\
-\kappa\log(2\sin2\phi) & \hphantom{\mbox{using the }}\mbox{new}\\
-\kappa\log2 & \hphantom{\mbox{using the }}\mbox{third}\,.
\end{cases}\label{eq:dD-long-string}
\end{align}
where $\kappa=\tfrac{1}{\pi}\log S$. The result for the physical
prescription is precisely that given in \cite{Forini:2012bb}. 

The result for the new sum prescription is not what we would expect
based on the giant magnon results, which is this:
\[
\delta\Delta=\delta\Delta_{\mathrm{phys}}+(c_{\mathrm{new}}-c_{\mathrm{phys}})\frac{\partial\Delta_{0}}{\partial h}=-\kappa\log2
\]
writing $\Delta_{0}=2h\log S$ for the leading term in \eqref{eq:spinning-Delta-minus-S}.
Note also that there is a divergence at $\phi=0$ i.e. at $\alpha=1$.
The reason for this is that $\omega_{5}^{B}$ and $\omega_{5}^{F}$
are declared massless in this limit, and hence omitted from \eqref{eq:dE-new-integral}.
Unlike the unambiguously massless modes $\omega_{1,2}^{B}$ and $\omega_{1,2}^{F}$
which clearly always cancel out, these do not. In fact the only difference
between the two sums (at $\alpha=1$) is that the new prescription
omits these two modes' contribution:
\begin{equation}
\sum_{n=-N_{1}}^{N_{1}}\omega_{5}^{B}-\omega_{5}^{F}=-\log N_{1}-\tfrac{1}{2}-\log2+\mathcal{O}\Big(\frac{1}{N_{1}}\Big).\label{eq:missing-modes-div}
\end{equation}

\subsection{Consequences}

In order to use these results to find $c$, we need to know the subleading
term in the expansion in $h$, i.e. $f_{1}$ in \eqref{eq:spinning-Delta-minus-S}:
\[
\delta\Delta=(2c+f_{1})\kappa\pi.
\]
It was noted in \cite{Forini:2012bb} that the Bethe equations at
$\alpha=1$ for the relevant $sl(2)$ sector are identical to those
for $AdS_{5}\times S^{5}$, and if one then assumes that the dressing
phase is also identical, then $f_{1}=\smash{-\frac{3\log2}{\pi}}$
as in \cite{Casteill:2007ct}. This led them to $c=\smash{\frac{\log2}{4\pi}}$.
Likewise at $\alpha=\tfrac{1}{2}$ the equations are the same as for
$AdS_{4}\times CP^{3}$, where the dressing phase was identical, and
$f_{1}=\smash{-\frac{3\log2}{2\pi}}$ \cite{Gromov:2008qe}; carrying
this over to the present $AdS_{3}$ gave $c=\smash{-\frac{\log2}{4\pi}}$.
We can now attempt a similar comparison for the new sum, where $\alpha=\smash{\frac{1}{2}}$
(and still $f_{1}=\smash{-\frac{3\log2}{2\pi}}$) leads to $c=+\frac{\log2}{2\pi}$. 

These values for $c$ are in all cases different to those from the
giant magnon, \eqref{eq:c-magnon}. It seems likely that what we are
learning here is that the dressing phase is \emph{not} the same as
it was for $AdS_{5}\times S^{5}$ and $AdS_{4}\times CP^{3}$. 

In this case we should instead use $c$ from the giant magnon and
$\delta\Delta$ from the spinning string to predict what $f_{1}$
will be. The answers are 
\[
f_{1}=\begin{cases}
-\frac{1}{\pi}\left[\log4+\sin^{2}\phi\log(\sin^{2}\phi)+\cos^{2}\phi\log(\cos^{2}\phi)\vphantom{1^{1}}\right] & \mbox{physical prescription}\\
-\frac{1}{\pi}\left[\log4+\log(2\sin2\phi)\vphantom{1^{1}}\right] & \mbox{new prescription}.
\end{cases}
\]
These two agree at $\phi=\frac{\pi}{4}$, but away from this they
disagree. Here is a graph: 
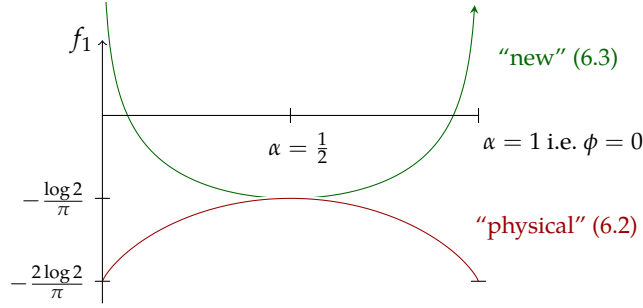
\[ 
\begin{tikzpicture} [scale=5]

\begin{scope}
\clip (-0.2,-0.5) rectangle (1.2,0.3);


\draw [-stealth, darkgreen] (0.005, 0.40278) -- (0.01, 0.293264) -- (0.015, 0.229538) -- (0.02, 0.184562) -- (0.025, 0.149862) -- (0.03, 0.121663) -- (0.035, 0.0979512) -- (0.04, 0.0775258) -- (0.045, 0.0596112) -- (0.05, 0.043678) -- (0.055, 0.0293488) -- (0.06, 0.0163448) -- (0.065, 0.00445443) -- (0.07, -0.00648684) -- (0.075, -0.0166094) -- (0.08, -0.0260184) -- (0.085, -0.0347998) -- (0.09, -0.0430247) -- (0.095, -0.0507529) -- (0.1, -0.0580348) -- (0.105, -0.0649133) -- (0.11, -0.0714256) -- (0.115, -0.0776036) -- (0.12, -0.0834755) -- (0.125, -0.0890656) -- (0.13, -0.0943957) -- (0.135, -0.099485) -- (0.14, -0.10435) -- (0.145, -0.109007) -- (0.15, -0.113469) -- (0.155, -0.117749) -- (0.16, -0.121858) -- (0.165, -0.125805) -- (0.17, -0.1296) -- (0.175, -0.133252) -- (0.18, -0.136768) -- (0.185, -0.140155) -- (0.19, -0.14342) -- (0.195, -0.146569) -- (0.2, -0.149607) -- (0.205, -0.152539) -- (0.21, -0.15537) -- (0.215, -0.158105) -- (0.22, -0.160746) -- (0.225, -0.1633) -- (0.23, -0.165768) -- (0.235, -0.168153) -- (0.24, -0.170461) -- (0.245, -0.172692) -- (0.25, -0.17485) -- (0.255, -0.176937) -- (0.26, -0.178955) -- (0.265, -0.180908) -- (0.27, -0.182797) -- (0.275, -0.184623) -- (0.28, -0.186389) -- (0.285, -0.188097) -- (0.29, -0.189748) -- (0.295, -0.191344) -- (0.3, -0.192886) -- (0.305, -0.194376) -- (0.31, -0.195815) -- (0.315, -0.197204) -- (0.32, -0.198545) -- (0.325, -0.199837) -- (0.33, -0.201084) -- (0.335, -0.202285) -- (0.34, -0.203442) -- (0.345, -0.204555) -- (0.35, -0.205626) -- (0.355, -0.206654) -- (0.36, -0.207642) -- (0.365, -0.208589) -- (0.37, -0.209496) -- (0.375, -0.210364) -- (0.38, -0.211194) -- (0.385, -0.211985) -- (0.39, -0.21274) -- (0.395, -0.213457) -- (0.4, -0.214139) -- (0.405, -0.214784) -- (0.41, -0.215394) -- (0.415, -0.215968) -- (0.42, -0.216508) -- (0.425, -0.217014) -- (0.43, -0.217485) -- (0.435, -0.217923) -- (0.44, -0.218327) -- (0.445, -0.218698) -- (0.45, -0.219036) -- (0.455, -0.219341) -- (0.46, -0.219614) -- (0.465, -0.219854) -- (0.47, -0.220062) -- (0.475, -0.220237) -- (0.48, -0.220381) -- (0.485, -0.220492) -- (0.49, -0.220572) -- (0.495, -0.22062) -- (0.5, -0.220636) -- (0.505, -0.22062) -- (0.51, -0.220572) -- (0.515, -0.220492) -- (0.52, -0.220381) -- (0.525, -0.220237) -- (0.53, -0.220062) -- (0.535, -0.219854) -- (0.54, -0.219614) -- (0.545, -0.219341) -- (0.55, -0.219036) -- (0.555, -0.218698) -- (0.56, -0.218327) -- (0.565, -0.217923) -- (0.57, -0.217485) -- (0.575, -0.217014) -- (0.58, -0.216508) -- (0.585, -0.215968) -- (0.59, -0.215394) -- (0.595, -0.214784) -- (0.6, -0.214139) -- (0.605, -0.213457) -- (0.61, -0.21274) -- (0.615, -0.211985) -- (0.62, -0.211194) -- (0.625, -0.210364) -- (0.63, -0.209496) -- (0.635, -0.208589) -- (0.64, -0.207642) -- (0.645, -0.206654) -- (0.65, -0.205626) -- (0.655, -0.204555) -- (0.66, -0.203442) -- (0.665, -0.202285) -- (0.67, -0.201084) -- (0.675, -0.199837) -- (0.68, -0.198545) -- (0.685, -0.197204) -- (0.69, -0.195815) -- (0.695, -0.194376) -- (0.7, -0.192886) -- (0.705, -0.191344) -- (0.71, -0.189748) -- (0.715, -0.188097) -- (0.72, -0.186389) -- (0.725, -0.184623) -- (0.73, -0.182797) -- (0.735, -0.180908) -- (0.74, -0.178955) -- (0.745, -0.176937) -- (0.75, -0.17485) -- (0.755, -0.172692) -- (0.76, -0.170461) -- (0.765, -0.168153) -- (0.77, -0.165768) -- (0.775, -0.1633) -- (0.78, -0.160746) -- (0.785, -0.158105) -- (0.79, -0.15537) -- (0.795, -0.152539) -- (0.8, -0.149607) -- (0.805, -0.146569) -- (0.81, -0.14342) -- (0.815, -0.140155) -- (0.82, -0.136768) -- (0.825, -0.133252) -- (0.83, -0.1296) -- (0.835, -0.125805) -- (0.84, -0.121858) -- (0.845, -0.117749) -- (0.85, -0.113469) -- (0.855, -0.109007) -- (0.86, -0.10435) -- (0.865, -0.099485) -- (0.87, -0.0943957) -- (0.875, -0.0890656) -- (0.88, -0.0834755) -- (0.885, -0.0776036) -- (0.89, -0.0714256) -- (0.895, -0.0649133) -- (0.9, -0.0580348) -- (0.905, -0.0507529) -- (0.91, -0.0430247) -- (0.915, -0.0347998) -- (0.92, -0.0260184) -- (0.925, -0.0166094) -- (0.93, -0.00648684) -- (0.935, 0.00445443) -- (0.94, 0.0163448) -- (0.945, 0.0293488) -- (0.95, 0.043678) -- (0.955, 0.0596112) -- (0.96, 0.0775258) -- (0.965, 0.0979512) -- (0.97, 0.121663) -- (0.975, 0.149862) -- (0.98, 0.184562) -- (0.985, 0.229538) -- (0.99, 0.293264); 

\draw [draw=white, thick, double=darkred, double distance=0.4pt] (0, -0.441271) -- (0.005, -0.431251) -- (0.01, -0.423445) -- (0.015, -0.41648) -- (0.02, -0.410064) -- (0.025, -0.404059) -- (0.03, -0.398381) -- (0.035, -0.392979) -- (0.04, -0.387813) -- (0.045, -0.382855) -- (0.05, -0.378082) -- (0.055, -0.373477) -- (0.06, -0.369025) -- (0.065, -0.364715) -- (0.07, -0.360535) -- (0.075, -0.356478) -- (0.08, -0.352536) -- (0.085, -0.348702) -- (0.09, -0.34497) -- (0.095, -0.341336) -- (0.1, -0.337794) -- (0.105, -0.334341) -- (0.11, -0.330972) -- (0.115, -0.327685) -- (0.12, -0.324475) -- (0.125, -0.321342) -- (0.13, -0.31828) -- (0.135, -0.31529) -- (0.14, -0.312367) -- (0.145, -0.309511) -- (0.15, -0.306719) -- (0.155, -0.303989) -- (0.16, -0.30132) -- (0.165, -0.29871) -- (0.17, -0.296158) -- (0.175, -0.293662) -- (0.18, -0.291222) -- (0.185, -0.288835) -- (0.19, -0.286502) -- (0.195, -0.28422) -- (0.2, -0.281988) -- (0.205, -0.279807) -- (0.21, -0.277674) -- (0.215, -0.275589) -- (0.22, -0.273551) -- (0.225, -0.27156) -- (0.23, -0.269614) -- (0.235, -0.267713) -- (0.24, -0.265857) -- (0.245, -0.264044) -- (0.25, -0.262274) -- (0.255, -0.260547) -- (0.26, -0.258862) -- (0.265, -0.257217) -- (0.27, -0.255614) -- (0.275, -0.254051) -- (0.28, -0.252528) -- (0.285, -0.251045) -- (0.29, -0.2496) -- (0.295, -0.248195) -- (0.3, -0.246827) -- (0.305, -0.245497) -- (0.31, -0.244205) -- (0.315, -0.24295) -- (0.32, -0.241732) -- (0.325, -0.240551) -- (0.33, -0.239406) -- (0.335, -0.238297) -- (0.34, -0.237223) -- (0.345, -0.236185) -- (0.35, -0.235183) -- (0.355, -0.234215) -- (0.36, -0.233282) -- (0.365, -0.232383) -- (0.37, -0.231519) -- (0.375, -0.230689) -- (0.38, -0.229893) -- (0.385, -0.229131) -- (0.39, -0.228402) -- (0.395, -0.227707) -- (0.4, -0.227045) -- (0.405, -0.226416) -- (0.41, -0.22582) -- (0.415, -0.225258) -- (0.42, -0.224728) -- (0.425, -0.22423) -- (0.43, -0.223765) -- (0.435, -0.223333) -- (0.44, -0.222933) -- (0.445, -0.222565) -- (0.45, -0.22223) -- (0.455, -0.221927) -- (0.46, -0.221655) -- (0.465, -0.221416) -- (0.47, -0.221209) -- (0.475, -0.221034) -- (0.48, -0.22089) -- (0.485, -0.220779) -- (0.49, -0.220699) -- (0.495, -0.220652) -- (0.5, -0.220636) -- (0.505, -0.220652) -- (0.51, -0.220699) -- (0.515, -0.220779) -- (0.52, -0.22089) -- (0.525, -0.221034) -- (0.53, -0.221209) -- (0.535, -0.221416) -- (0.54, -0.221655) -- (0.545, -0.221927) -- (0.55, -0.22223) -- (0.555, -0.222565) -- (0.56, -0.222933) -- (0.565, -0.223333) -- (0.57, -0.223765) -- (0.575, -0.22423) -- (0.58, -0.224728) -- (0.585, -0.225258) -- (0.59, -0.22582) -- (0.595, -0.226416) -- (0.6, -0.227045) -- (0.605, -0.227707) -- (0.61, -0.228402) -- (0.615, -0.229131) -- (0.62, -0.229893) -- (0.625, -0.230689) -- (0.63, -0.231519) -- (0.635, -0.232383) -- (0.64, -0.233282) -- (0.645, -0.234215) -- (0.65, -0.235183) -- (0.655, -0.236185) -- (0.66, -0.237223) -- (0.665, -0.238297) -- (0.67, -0.239406) -- (0.675, -0.240551) -- (0.68, -0.241732) -- (0.685, -0.24295) -- (0.69, -0.244205) -- (0.695, -0.245497) -- (0.7, -0.246827) -- (0.705, -0.248195) -- (0.71, -0.2496) -- (0.715, -0.251045) -- (0.72, -0.252528) -- (0.725, -0.254051) -- (0.73, -0.255614) -- (0.735, -0.257217) -- (0.74, -0.258862) -- (0.745, -0.260547) -- (0.75, -0.262274) -- (0.755, -0.264044) -- (0.76, -0.265857) -- (0.765, -0.267713) -- (0.77, -0.269614) -- (0.775, -0.27156) -- (0.78, -0.273551) -- (0.785, -0.275589) -- (0.79, -0.277674) -- (0.795, -0.279807) -- (0.8, -0.281988) -- (0.805, -0.28422) -- (0.81, -0.286502) -- (0.815, -0.288835) -- (0.82, -0.291222) -- (0.825, -0.293662) -- (0.83, -0.296158) -- (0.835, -0.29871) -- (0.84, -0.30132) -- (0.845, -0.303989) -- (0.85, -0.306719) -- (0.855, -0.309511) -- (0.86, -0.312367) -- (0.865, -0.31529) -- (0.87, -0.31828) -- (0.875, -0.321342) -- (0.88, -0.324475) -- (0.885, -0.327685) -- (0.89, -0.330972) -- (0.895, -0.334341) -- (0.9, -0.337794) -- (0.905, -0.341336) -- (0.91, -0.34497) -- (0.915, -0.348702) -- (0.92, -0.352536) -- (0.925, -0.356478) -- (0.93, -0.360535) -- (0.935, -0.364715) -- (0.94, -0.369025) -- (0.945, -0.373477) -- (0.95, -0.378082) -- (0.955, -0.382855) -- (0.96, -0.387813) -- (0.965, -0.392979) -- (0.97, -0.398381) -- (0.975, -0.404059) -- (0.98, -0.410064) -- (0.985, -0.41648) -- (0.99, -0.423445) -- (0.995, -0.431251) -- (1, -0.441271);

\end{scope};

\draw (0,0) -- (1,0) ;
\draw [->] (0,-0.5) -- (0,0.2) node [anchor=east] {$f_1$};
\draw (1,0.2mm) -- (1,-0.2mm) node[anchor=north west, xshift=-3mm] {\small $\alpha=1$ i.e. $\phi=0$}; 

\draw (0.5,0.2mm) -- (0.5,-0.2mm) node[anchor=north] {\small $\alpha=\tfrac{1}{2}$}; 

\node[darkgreen, anchor=west] at (0.98,0.15) {\small ``new'' (\ref*{eq:f1-new}) };
\node[darkred, anchor=west] at (0.92,-0.3) {\small ``physical'' \eqref{eq:f1-phys}};

\draw (0.2mm,-0.220636) -- (-0.2mm,-0.220636) node [anchor=east] {\small $-\frac{\log 2}{\pi}$};

\draw (0.2mm,-0.441271) -- (-0.2mm,-0.441271) node [anchor=east] {\small $-\frac{2 \log 2}{\pi}$};
\draw (10.2mm,-0.441271) -- (9.8mm,-0.441271);

\end{tikzpicture}
\] This difference is quite unlike anything seen in the $AdS_{4}\times CP^{3}$
case: The one-loop (i.e. $1/g$) corrections to the spinning string
obtained using these two prescriptions cannot both follow from the
same function of $h$. Thus we must conclude that at least one of
them is incorrect. Given the divergence seen above at $\phi=0$, it
seems reasonable to say that it is the new prescription which is at
fault. 

Finally note that for the third cutoff \eqref{eq:third-way}, results
both for the magnon and for the spinning string differ from those
for the physical cutoff only by a $\log2$ term. These cancel out
to give exactly the same prediction for $f_{1}$ i.e. \eqref{eq:f1-phys}.

\section{Comments and Conclusions\label{sec:Comments-and-Conclusions}}

The first paper to write down the subleading term of the interpolating
function $h(\lambda)$ for $\adssss$ was \cite{Forini:2012bb}, who
gave 
\[
c=\begin{cases}
-\frac{\log2}{4\pi} & \alpha=\frac{1}{2}\\
\frac{\log2}{4\pi} & \alpha=1
\end{cases},\quad\mbox{assuming \quad}f_{1}=\begin{cases}
-\frac{3\log2}{2\pi} & \mbox{ as for }CP^{3}\\
-\frac{3\log2}{\pi} & \mbox{ as for }S^{5}.
\end{cases}
\]
These values for $f_{1}$ come from \cite{Forini:2012bb}'s observation
that the Bethe equations for this $sl(2)$ sector are the same as
those for $AdS_{4}\times CP^{3}$ or $AdS_{5}\times S^{5}$ at these
two values of $\alpha$, and then tentatively assuming that the dressing
phase is also the same as that for both previous correspondences.
As they note, there is no particular reason to think that this is
true, and comparison with the giant magnon results now shows it not
to be so. For the same value of $\alpha$, the magnons give 
\begin{equation}
c=\begin{cases}
-\frac{\log2}{2\pi} & \alpha=\frac{1}{2}\\
0 & \alpha=1.
\end{cases}\qquad\qquad\label{eq:two-c-magnon}
\end{equation}
Turning this comparison around we can instead use the magnon calculation
\eqref{eq:c-magnon} to predict the one-loop correction which should
arise from the correct dressing phase, for all $\alpha$. This gives:
\begin{equation}
f_{1}=-\frac{1}{\pi}\left[2\log2+\alpha\log\alpha+(1-\alpha)\log(1-\alpha)\vphantom{1^{1}}\right].\label{eq:f1-phys}
\end{equation}
The giant magnon calculation here uses the algebraic curve and thus
omits massless modes. This is a possible source of error, but see
points \ref{enu:Per-light}, \ref{enu:per-heavy} below for evidence
against this. The spinning string calculation includes the massless
modes, but they cancel among themselves. 

This result is for the ``physical'' summation prescription, i.e.
using a cutoff at the same energy (or mode number) for all modes.
One can instead consider the ``new'' prescription, defined with
a cutoff at fixed radius in the spectral plane. This is one possible
generalisation of the cutoff introduced by \cite{Gromov:2008fy} for
$\adscp$, but the story is a little different here:
\begin{enumerate}[resume]
\item If this prescription is equally valid, then adopting it should affect
all one-loop results in the same way, namely $\delta E_{\mathrm{new}}-\delta E_{\mathrm{phys}}=(c_{\mathrm{new}}-c_{\mathrm{phys}})\partial E_{0}/\partial h\smash{\big\vert_{h=2g}}$.
However this is not what happens. One way to say this is to use $c=0$
from the giant magnon and $\delta E$ from the spinning string to
predict $f_{1}$. This gives 
\begin{equation}
f_{1}=-\frac{1}{\pi}\left[2\log2+\tfrac{1}{2}\log\alpha+\tfrac{1}{2}\log(1-\alpha)\vphantom{1^{1}}\right]\label{eq:f1-new}
\end{equation}
disagreeing with \eqref{eq:f1-phys} at $\alpha\neq\tfrac{1}{2}$.
However $f_{1}$ is part of an expansion in $h$ which should be independent
of the prescription. \label{enu:new-f1-diff}
\item That there is something wrong with the new prescription is most clear
at $\alpha=1$. Here one of the modes it deems massless in the limit
in fact plays an important role (for the long spinning string), and
the effect of changing to the new prescription is simply to remove
this from the sum, \eqref{eq:missing-modes-div}, producing a divergence.
We would very much like to have a smooth $\alpha\to1$ limit. \label{enu:new-gives-div}
\item Instead of the new prescription we can consider a third prescription,
\eqref{eq:third-way}, cutting off both kinds of light modes at half
the energy of the heavy modes. At $\alpha=\tfrac{1}{2}$, and in $AdS_{4}\times CP^{3}$,
this is identical. In some way it may be closer to the spirit of \cite{Gromov:2008fy},
in that it treats heavy modes with mode number $2n$ alongside light
modes with mode number $n$. However this cutoff seems unnatural since
at $\alpha=1$ it doesn't treat all of the massive modes on an equal
footing. It leads to the same $f_{1}$ as the physical prescription.
\label{enu:concl-third-way}
\end{enumerate}
In the appendices I also work out some finite-$J$ corrections to
giant magnons, since this is easily done with the algebraic curve.
The main points to note are:
\begin{enumerate}[resume]
\item The exponent of the classical $\mu$-term \eqref{eq:dE-mu-final}
depends on the ``mass'' of the giant magnon. This is just a result
of embedding the well-known solutions into this spacetime, \eqref{eq:X-embed}. 
\item The exponents of the one-loop F-terms depend on the masses of the
virtual particles. This can be thought of as being a consequence of
the scaling of the AFS phase introduced by \cite{OhlssonSax:2011ms},
which itself may be thought of as a result of the scaling of the time
delay when magnon scattering happens on a sphere of smaller radius.
Is has the effect that, unlike the $AdS_{4}\times CP^{3}$ case, in
general the bound-state and twice-wrapped contributions will not coincide
\cite{Abbott:2011tp}. Since each F-term depends only on one mass
of virtual particle, these terms are unaffected by the choice of regularisation
prescription \cite{Ahn:2010eg,Abbott:2010yb}. 
\item Taking the $\alpha\to1$ limit gives the same F-term corrections as
those calculated directly from the algebraic curve for $AdS_{3}\times S^{3}$,
\eqref{eq:T4-F-left}. This limit removes the trivial terms (corresponding
to an S-matrix element of 1) arising from the fact that giant magnons
on different spheres pass each other on the worldsheet without interacting. 
\end{enumerate}

\subsection{Relation to other work}

A recent paper by Sundin and Wulff \cite{Sundin:2012gc} calculates,
among other things, one-loop mass corrections to string states in
the near-BMN limit of $AdS_{3}\times S^{3}\times S^{3}\times S^{1}$.
It is interesting to compare results since their paper works from
the full Green--Schwartz action and thus includes all of the massless
modes. 
\begin{enumerate}[resume]
\item For the light bosons these mass corrections should agree with the
small-$p$ limit of the corrections to giant magnons \cite{Klose:2007rz},
and indeed they match perfectly for both physical (``WS'') and new
(``AC'') prescriptions, \eqref{eq:two-c-magnon}. These are given
for $\alpha=\tfrac{1}{2}$ and $\alpha=1$. \label{enu:Per-light}
\item The correction for the heavy boson is computed by \cite{Sundin:2012gc}
at all $\alpha$, and while using the physical sum  gives a result
consistent with the giant magnon's \eqref{eq:c-magnon}, using the
new sum gives $c=\tfrac{1}{2\pi}\sin^{2}2\phi\,\log(\sin^{2}2\phi)$
rather than zero.%
\footnote{While this mode is not a cromulent giant magnon, since it is in AdS,
one can nevertheless attempt a \naive treatment of a giant ``4''
mode by turning on $G_{1}=\smash{\frac{-1}{2\sin^{2}\phi}}G_{\mathrm{mag}}$,
$G_{2}=-G_{\mathrm{mag}}$ and $G_{3}=\smash{\frac{-1}{2\cos^{2}\phi}}G_{\mathrm{mag}}$.
Then the calculation of $\delta E$ looks very much the same as that
in section \ref{sec:Giant-Magnons} above, and in particular is zero
for the new sum.\label{fn:non-cromulent}%
} This seems strange as $c$ should be universal. \medskip \\
Note however that (as in \cite{Abbott:2011xp}) the new prescription
can only be implemented for the tadpole diagrams. For the bubbles
the same loop momentum applies to two modes of different masses, in
this case one of each mass of light mode. (Thus there is no issue
at $\phi=\tfrac{\pi}{4}$, where this calculation gives $c=0$.) The
massless modes play no role in this calculation.\label{enu:per-heavy}
\item Virtual massless modes \emph{do} appear to play a role for the corrections
to the light bosons. However let me observe that simply deleting all
diagrams containing them%
\footnote{That is, delete all integrals $I_{n}^{s}(\ldots m_{4})$ in $\mathcal{A}_{B}^{2}$
above (4.22) and $I_{n}^{s}(m_{4})$ in $\mathcal{A}_{T}^{i}$ (4.23),
where $m_{4}\ll1$ is \cite{Sundin:2012gc}'s IR regulator for the
massless modes. However for this to work we must use the canonical
dimensionally regulated integrals with measure $\smash{\int d^{d}k}/(2\pi)^{d}$
\cite{Peskin:1995ev}, rather than (4.10)'s $\smash{\int d^{d}k}/(2\pi)^{2}$;
this does not seem to affect final results such as (4.25). %
} does not change the result. This is an extremely \naive thing to
do (since for instance the cubic interaction $\mathcal{L}_{3}$, \cite{Sundin:2012gc}'s
equation (3.3), has terms linear in the massless boson) but nevertheless
perhaps interesting.%
\footnote{I am grateful to Per Sundin for discussions on this point. %
} 
\end{enumerate}
There are some points of overlap with other papers studying integrability
in $AdS_{3}\times S^{3}\times S^{3}\times S^{1}$ worth mentioning:
\begin{enumerate}[resume]
\item For the case $\alpha=1$ (i.e. $AdS_{3}\times S^{3}$) the energy
corrections for giant magnons here agree with those of \cite{David:2010yg},
who also used the algebraic curve formalism. In particular we agree
that $c=0$ in this limit.
\item As mentioned briefly in the introduction the  recent papers \cite{Ahn:2012hw}
and \cite{Borsato:2012ud} (which appeared while this was in preparation)
each propose an S-matrix for this system. This should agree with terms
in the F-term corrections calculated here, and a quick comparison
shows agreement up to certain phases. See discussion in appendix \ref{sec:Finite-Size}. 
\end{enumerate}
Finally, some comments on the closely related issues in $AdS_{4}\times CP^{3}$.
There, in all calculations to date one can use either the physical
or the new prescription, and the change in the results is always equivalent
to\vspace{-2mm} 
\[
c_{\mathrm{phys}}=-\frac{\log2}{2\pi},\qquad c_{\mathrm{new}}=0
\]
without running into the problems of points \ref{enu:new-f1-diff},
\ref{enu:new-gives-div} above. Nevertheless various arguments have
been advanced for choosing one or the other. Below is a very brief
list of these; most will also apply to the present $AdS_{3}\times S^{3}\times S^{3}\times S^{1}$
case.
\begin{enumerate}[resume]
\item In favour of the physical sum, in \cite{Abbott:2011xp} we pointed out
that it is difficult to see how to implement the new sum for Feynman
diagrams containing modes of different masses in the same loop. (See
also point \ref{enu:per-heavy}.) More strongly, \cite{LopezArcos:2012gb}
uses a general condition from \cite{Nastase:1998sy} that the vacuum
energy should not depend on the topology when the soliton mass is
zero. It would be interesting to see what this says about the present
case. 
\item In favour of the new sum, \cite{Astolfi:2011ju,Astolfi:2011bg} argue
that unitarity requires that the energy of two light modes near to
their cutoff should correspond to that of a heavy mode near its cutoff.
(See also point \ref{enu:concl-third-way}.) Perhaps also under this
heading it should be mentioned that \cite{Leoni:2010tb,Minahan:2009wg,Minahan:2009aq}
conjecture an all-loop $h(\lambda)$ consistent with their 4-loop
weak-coupling result; this gives $c=0$. And lastly, \cite{Zarembo:2009au}
might also be included (see also \cite{Sundin:2009zu,Abbott:2011xp,Sundin:2012gc})
on the grounds that questions of the position of the heavy mode's
pole vs. the two-particle cut are only subtle when $c=0$. \label{enu:the-last-comment}
\end{enumerate}

\subsection*{Acknowledgements}

I would like to thank Diego Bombardelli, Justin David, Olof Ohlsson
Sax, Bogdan Stefa\'{n}ski, Per Sundin and Kostya Zarembo for discussions
and correspondence, and IFT-UNESP S\~{a}o Paulo for hospitality while
finishing this. 

\appendix

\section{Finite-$J$ Corrections\label{sec:Finite-Size}}

While this is somewhat aside from the paper, the algebraic curve can
also be used to compute certain finite-size corrections. The classical
$\mu$-term is mostly just a check that things are working well. The
one-loop F-terms will perhaps teach us something about the $\alpha\to1$
limit.

\subsection{Classical $\mu$-term}

The first class of finite-$J$ corrections are $\mu$-terms, suppressed
by $e^{-m_{r}\Delta/E}$. The leading $\mu$-term is the classical
correction away from $J=\infty$, and from the sigma-model point of
view we expect to get results almost identical to those for magnons
in $\mathbb{R}\times S^{2}$ \cite{Arutyunov:2006gs,Astolfi:2007uz}
or dyonic magnons in $\mathbb{R}\times S^{3}$ \cite{Hatsuda:2008gd,Minahan:2008re,Okamura:2006zv}
since the same string solutions can be embedded into this spacetime.%
\footnote{By contrast, in $AdS_{4}\times CP^{3}$ only the non-dyonic solutions
are the same \cite{Grignani:2008te,Lee:2008ui,Abbott:2008qd}, the
dyonic solutions are new \cite{Abbott:2009um,Hollowood:2009sc} and
known at finite $J$ only in the algebraic curve language.%
} Nevertheless it is a check of the algebraic curve description. 

\usetikzlibrary{snakes}

To work this out we must use a different algebraic curve solution,
and following \cite{Minahan:2008re,Lukowski:2008eq,Sax:2008in,Abbott:2009um}
we can use the following (approximate) form for the magnon resolvent:
\begin{equation}
G_{\mathrm{finite}}(x)=-2i\log\Bigg(\frac{\sqrt{x-a}+\sqrt{x-b}}{\sqrt{x-\bar{a}}+\sqrt{x-\bar{b}}}\Bigg)\label{eq:def-Gfinite}
\end{equation}
where the branch points are 
\[
a=\Xp(1+\tfrac{\delta}{2}e^{i\psi}),\qquad b=\Xp(1-\tfrac{\delta}{2}e^{i\psi})
\]
and complex conjugates (with $\delta$ real). Recall that this arises
from the two-cut solution, reorganising the two square root cuts (drawn~$\tikz [baseline=-0.3em] \draw [darkred, semithick, double, cap=rect] (0.7,0) to [in=10,out=170] (0,0);$)
at the cost of producing a log cut (drawn~$\tikz [baseline=-0.3em] \draw [darkgreen, semithick, snake=coil, segment aspect=0, segment amplitude=1] (0,0) -- (0.7,0);$)
like this: 
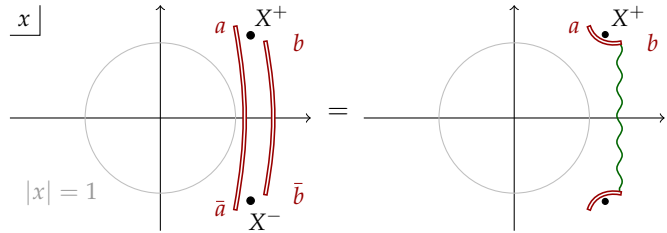
\[
\begin{tikzpicture}[scale=1, baseline] 

\draw [->] (-2,0) -- (2,0); 
\draw [->] (0,-1.5) -- (0,1.5); 
\draw (-2,1.1) -- (-1.6,1.1) -- (-1.6,1.5); 
\node [anchor=south east] at (-1.6,1.1) {\small $x$};
\draw [lightgray] (0,0) circle (1);
\node [anchor=north east, gray!80] at (-0.71,-0.71) {\small $\lvert x \rvert = 1$};

\draw [darkred, semithick, double, cap=rect] (1,-1.2) to [out=80,in=280] (1,1.2);
\draw [darkred, semithick, double, cap=rect] (1.4,-1) to [out=80,in=280] (1.4,1);

\draw [fill] (1.2,1.1) circle (0.05);
\node [anchor=east, darkred] at (1,1.2) {\small $a$};
\node [anchor=south] at (1.2,1.1) {\small $\ \ \ \ \Xp$};
\node [anchor=west, darkred] at (1.4,1) {\small $b$};

\draw [fill] (1.2,-1.1) circle (0.05);
\node [anchor=east, darkred] at (1,-1.2) {\small $\bar{a}$};
\node [anchor=north] at (1.2,-1.1) {\small $\ \ \Xm$};
\node [anchor=west, darkred] at (1.4,-1) {\small $\bar{b}$};

\end{tikzpicture}
\ = \ 
\begin{tikzpicture}[scale=1, baseline] 
\draw [->] (-2,0) -- (2,0); 
\draw [->] (0,-1.5) -- (0,1.5); 
\draw [lightgray] (0,0) circle (1);

\draw [fill] (1.21,1.11) circle (0.04);
\draw [fill] (1.21,-1.11) circle (0.04);

\draw [darkgreen, semithick, snake=coil, segment aspect=0, segment amplitude=1] (1.4,1) -- (1.4,-1);

\draw [darkred, semithick, double, cap=rect] (1,1.2) to [out=290, in=190] (1.4,1);
\draw [darkred, semithick, double, cap=rect] (1,-1.2) to [out=70, in=170] (1.4,-1);

\node [anchor=east, darkred] at (1,1.2) {\small $a$};
\node [anchor=south] at (1.2,1.1) {\small $\ \ \ \ \ \Xp$};
\node [anchor=west, darkred] at (1.4,1) {\small $b$};

\end{tikzpicture}
\]  Thus $\delta\to0$ gives us the $J=\infty$ giant magnon \eqref{eq:G1-Gmag}
above \cite{Minahan:2006bd,Vicedo:2007rp,Chen:2007vs}. 

To use this resolvent here, we should set $G_{1}(x)=\smash{\frac{1}{2\sin^{2}\phi}}[G_{\mathrm{finite}}(x)-\tfrac{1}{2}G_{\mathrm{finite}}(0)]$
in \eqref{eq:quasi-q}. Expanding the charges $J'$, $Q$, and $P$
in $\delta$, the correction to the dispersion relation is obtained
as follows: 
\begin{align*}
\delta E^{\mu} & =\Delta-J'\:-\sqrt{Q{}^{2}\sin^{4}\phi+16g^{2}\sin^{2}\frac{P}{2}}\\
 & =\delta^{2}\,\frac{g}{4}\,\cos(2\psi)\,\sin\frac{p}{2}+\mathcal{O}(\delta^{4})\qquad\mbox{(non-dyonic).}
\end{align*}
The factor $\cos(2\psi)$ gives the effect of the angle between subsequent
magnons, as we discussed in \cite{Abbott:2009um}. Next we must fix
$\delta$ by demanding matching across the cut. To do this, use \eqref{eq:jump-condition-p}
in the form:
\begin{align}
p_{1}(\Xp-i0) & =p_{1}(\Xp+i0)-A_{1\ell}p_{\ell}(\Xp+i0)+2\pi n\label{eq:mu-term-p-matching}\\
\shortintertext{i.e.}G_{\mathrm{finite}}(\Xp-i0) & =(1-4\sin^{2}\phi)\: G_{\mathrm{finite}}(\Xp)-4\sin^{4}\phi\frac{\Delta}{2g}\frac{\Xp}{\Xp^{2}-1}-2\sin^{2}\phi\:2\pi n.\nonumber 
\end{align}
Expanding and solving then gives 
\begin{align*}
\delta^{2} & =-16(\Xp-\Xm)^{2}\exp-i\Big(\frac{\Delta\sin^{2}\phi}{g}\frac{\Xp}{\Xp^{2}-1}+2\pi n+2\psi+\frac{\pi}{2\sin^{2}\phi}\Big)\\
 & =64\,\exp\Big(\frac{-\Delta\sin^{2}\phi}{2g\,\sin^{2}\frac{p}{2}}\Big)\,\sin^{2}\frac{p}{2}\qquad\mbox{(non-dyonic).}
\end{align*}
The second line assumes that $\delta$ is real, which imposes $p+2n\pi+2\psi+\frac{1}{2}\pi\cosec^{2}\phi=0$.
Then setting $\psi=\frac{\pi}{2}$ the correction is 
\begin{equation}
\delta E^{\mu}=-16g\, e^{-\Delta\sin^{2}\phi\big/2g\sin^{2}\frac{p}{2}}\sin^{3}\frac{p}{2}.\label{eq:dE-mu-final}
\end{equation}
This $\sin^{3}\frac{p}{2}$ behaviour matches what was given by AFZ
\cite{Arutyunov:2006gs}; the scaling of the exponent comes from placing
this into a spheres of a different radius --- see appendix \ref{sec:Worldsheet-Giant-Magnons}.

\subsection{One-loop F-terms}

The second class of finite-$J$ corrections are F-terms, suppressed
by both $e^{-m_{r}\Delta/g}$ and $1/g$ relative to the leading term
$E_{0}$. These can readily be computed using the one-loop mode sum
as above: they are simply the subsequent terms in the expansion of
the cotangent in \eqref{eq:dE-with-cot}, for which the integral can
be evaluated using the saddle point at $x=i$. Similar calculations
were done by \cite{Gromov:2008ie} in $S^{5}$ and \cite{Shenderovich:2008bs,Abbott:2010yb,Ahn:2010eg}
in $CP^{3}$. 

The expansion is as follows: 
\begin{align*}
\cot(\pi n)\,\pi n' & =\partial_{x}\log\big(\sin(\pi n)\big)\\
 & =\pm i\pi n'+\partial_{x}\log\big(1-e^{\mp i\,2\pi n}\big)\displaybreak[0]\\
 & =\pm i\left[\pi n'+e^{\mp i\,2\pi n}2\pi n'+\frac{e^{\mp2i\,2\pi n}}{2}2\pi n'+\frac{e^{\mp3i\,2\pi n}}{3}2\pi n'+\ldots\right].
\end{align*}
In \eqref{eq:dE-with-cot} above, the first term gave the infinite-volume
one-loop correction. Subsequent terms can be written 
\begin{align*}
\delta E^{F} & =\frac{-i}{2\pi}\int_{U_{+}}dx\sum_{r}(-1)^{F_{r}}\left[e^{-i2\pi n_{r}(x)}+\frac{e^{-2i\,2\pi n_{r}(x)}}{2}+\frac{e^{-3i\,2\pi n_{r}(x)}}{3}+\ldots\right]\partial_{x}\Omega_{r}(x)\\
 & =\sum_{\ell=1,2,3,\ldots}\frac{1}{\ell}\sum_{r=1,3,4.}\sqrt{\frac{m_{r}}{2\pi\kappa}}F_{m_{r}}^{(\ell)}(i)
\end{align*}
where we use the saddle point at $x=i$, and define $F_{m}^{(\ell)}(x)=\sum_{r:m_{r}=m}(-1)^{F_{r}}\exp(-i\,\ell\,2\pi n_{r}(x)\,)$.
This is the factor which in the L\"{u}scher formula is $\sum_{b}(S_{b1}^{b1})^{\ell}$
\cite{Janik:2007wt,Heller:2008at}. (The $\Omega'$ factor is the
Jacobian there.) 

Here are some of the resulting integrands, taking as the classical
solution the magnon in $G_{3}$ (i.e. the magnon in the sphere which
survives at $\phi=0$). First from the light modes $1,1f,\bar{1},\bar{1}f$
(in that order) writing just the $\ell=1$ case:
\begin{equation}
F_{m_{1}}^{(1)}=\exp\negmedspace\Big(-\frac{\Delta\sin^{2}\phi}{g}\frac{i\, x}{x^{2}-1}\Big)\left[1-\sqrt{\frac{\Xm}{\Xp}}\frac{x-\Xp}{x-\Xm}+1-\sqrt{\frac{\Xp}{\Xm}}\frac{1-x\Xm}{1-x\Xp}\right].\label{eq:F-diff-light}
\end{equation}
 Next from the corresponding modes of mass $\cos^{2}\phi$, the same
as the giant magnon, namely $3,3f,\bar{3},\bar{3}f$ 
\begin{equation}
\negthickspace F_{m_{3}}^{(1)}=e^{-\frac{\Delta\cos^{2}\phi}{g}\frac{i\, x}{x^{2}-1}}\negmedspace\left[\frac{\Xp}{\Xm}\negmedspace\left(\frac{x-\Xm}{x-\Xp}\right)^{2}-\sqrt{\frac{\Xp}{\Xm}}\frac{x-\Xm}{x-\Xp}+\frac{\Xm}{\Xp}\negmedspace\left(\frac{1-x\Xp}{1-x\Xm}\right)^{2}-\sqrt{\frac{\Xm}{\Xp}}\frac{1-x\Xp}{1-x\Xm}\right].\label{eq:F-same-same}
\end{equation}
And finally from the heavy modes $4,4f,\bar{4},\bar{4}f$:
\begin{equation}
F_{m_{4}}^{(1)}=e^{-\frac{\Delta}{g}\frac{i\, x}{x^{2}-1}}\left[1-\sqrt{\frac{\Xp}{\Xm}}\frac{x-\Xm}{x-\Xp}+1-\sqrt{\frac{\Xm}{\Xp}}\frac{1-x\Xp}{1-x\Xm}\right].\label{eq:F-heavy}
\end{equation}
Two comments on these:
\begin{itemize}
\item Note that as in $AdS_{4}\times CP^{3}$ the modes of different masses
lead to different factors in the exponential, but now there are three
terms. These are separately finite and thus the regulator used is
irrelevant. These exponential factors can perhaps be thought of as
coming from the scaled $\sigma_{\mathrm{AFS}}$ as in (2.11) of \cite{OhlssonSax:2011ms}. 
\item About the heavy modes, note also that their terms here are products
of the constituent light modes', $4=1f+3f$ etc. This is what we would
expect if the S-matrices for these can be made by fusion, as in \cite{Abbott:2011tp}.
Unlike $AdS_{4}\times CP^{3}$, the $\ell=2$ term from a light mode
wrapping twice will not coincide with the $\ell=1$ heavy mode \cite{Abbott:2011tp}.
\end{itemize}

\subsection{Comparison with S-matrices}

The F-term formulae above should agree with certain diagonal elements
of the recently published S-matrices for this system. This section
aims to make some quick comparisons.%
\footnote{I am grateful to Diego Bombardelli and Olof Ohlsson Sax for discussing
their results.%
}

Comparing first to Ahn and Bombardelli's \cite{Ahn:2012hw}, for the
modes on the left the terms of their equation (2.14) we need are 
\begin{align*}
S^{(33)}(p_{1},p_{2}) & =S_{0}(p_{1},p_{2})\hat{S}(p_{1},p_{2})\\
S^{(13)}(p_{1},p_{2}) & =\hat{S}(p_{1},p_{2})
\end{align*}
where $x_{p_{1}}^{\pm}\approx x$ for the virtual particle,  $x_{p_{2}}^{\pm}=\Xpm$
for the physical giant magnon, and the matrix part is thus 
\[
\hat{S}=\begin{cases}
1, & \mbox{bose-bose}\\
\dfrac{x-\Xp}{x-\Xm}, & \mbox{fermi-bose}.
\end{cases}
\]
For the scalar factor $S_{0}$, only the classical AFS \cite{Arutyunov:2004vx}
term in the BES phase \cite{Beisert:2006ez} matters, giving 
\[
S_{0}=\sigma_{\mathrm{AFS}}^{2}(p_{1},p_{2})\: e^{-ip_{1}+ip_{2}}=\left(\frac{x-1/\Xp}{x-1/\Xm}\right)^{2}\frac{\Xp}{\Xm}e^{-2i\frac{x}{x^{2}-1}E/h}+\mathcal{O}\Big(\frac{1}{h}\Big).
\]
Then  since $X^{+}=1/X^{-}+\mathcal{O}(1/g)$ we have agreement with
the first two terms of \eqref{eq:F-diff-light} and of \eqref{eq:F-same-same},
apart from phases $e^{-ip_{2}/2}=\sqrt{\Xm/\Xp}$ for the fermions
($1f$ and $3f$) which will arise from going to the string frame. 

Note however that I have ignored here the phase $e^{-2i\frac{x}{x^{2}-1}E/h}$
which is part of $\sigma_{\mathrm{AFS}}$. In the more familiar $AdS_{5}$
and $AdS_{4}$ cases the same power of $\sigma_{\mathrm{AFS}}$ appears
in all terms, and changes the exponent $e^{-iq_{\star}L}$ from the
L\"{u}scher formula (with $L=J$) into $e^{-\Delta/h}$ from the
algebraic curve. The absence of such a phase for $S^{(13)}$ is a
feature of \cite{Ahn:2012hw}'s S-matrix designed to match the Bethe
equations of \cite{OhlssonSax:2011ms}. 

For the modes on the right, we need to look at 
\begin{align*}
S^{(\bar{3}3)}(p_{1},p_{2}) & =\widetilde{S}_{0}(p_{1},p_{2})\hat{S}(p_{1},p_{2})\\
S^{(\bar{1}3)}(p_{1},p_{2}) & =\hat{S}(p_{1},p_{2}).
\end{align*}
Here $\widetilde{S}_{0}(p_{1},p_{2})=\sigma_{\mathrm{AFS}}^{-2}(p_{1},\bar{p}_{2})\, e^{ip_{1}+ip_{2}}$
where the bar means $x_{\bar{p}}^{\pm}=1/x_{p}^{\pm}$, and $\hat{S}$
is as before. These match the last two terms of \eqref{eq:F-diff-light}
and \eqref{eq:F-same-same}, up to the same two phase issues as for
the left-hand modes. 

Next consider the S-matrix given by Borsato, Ohlsson Sax and Sfondrini
in \cite{Borsato:2012ud}. This is written in terms of four unfixed
phases, related by crossing relations. The coefficients relevant for
\eqref{eq:F-same-same} are $\mathsf{A}_{pq}^{LL}$, $\mathsf{B}_{pq}^{LL}$,
$\mathsf{A}_{pq}^{LR}$, $\mathsf{C}_{pq}^{LR}$, and we should use
the string frame expressions in appendix E. Then setting $x_{q}^{\pm}=x+\mathcal{O}(1/g)$
for the virtual particle, and $x_{p}^{\pm}=X^{\pm}$ for the physical
giant magnon, the unfixed phases are 
\[
S_{pq}^{LL}=\sqrt{\frac{x_{p}^{+}}{x_{p}^{-}}}\:\frac{x-x_{p}^{-}}{x-x_{p}^{+}}\:\sigma_{3}(x,x_{p}^{\pm}),\qquad\tau_{pq}^{LR}=\sqrt{\frac{x_{p}^{+}}{x_{p}^{-}}}\:\frac{x-1/x_{p}^{+}}{x-1/x_{p}^{-}}\:\sigma_{3}(x,x_{p}^{\pm}).
\]
Here some phase $\smash{\sigma_{3}}$ is needed because the L\"{u}scher
formula gives $e^{-i\, m_{r}q_{\star}L}$ (with $q_{\star}=\smash{\frac{1}{h}\frac{x}{x^{2}-1}}$)
rather than the first factor in \eqref{eq:F-same-same}. If $L=\smash{J'}$
then this could be provided (in this limit) by some power of the AFS
phase. 

In order to check crossing symmetry we need first (5.27): 
\[
S_{pq}^{LR}=\frac{1}{\zeta_{pq}}\tau_{pq}^{LR}=\sqrt{\frac{x_{p}^{+}}{x_{p}^{-}}}\left(\frac{x-1/x_{p}^{+}}{x-1/x_{p}^{-}}\right)^{3/2}\sigma_{3}.
\]
 Then using $x_{\bar{q}}^{\pm}=1/x_{q}^{\pm}\approx1/x$, we obtain
\[
S_{pq}^{LL}S_{p\bar{q}}^{LR}=\sqrt{\frac{x_{p}^{-}}{x_{p}^{+}}}\sqrt{\frac{x-x_{p}^{+}}{x-x_{p}^{-}}}\sigma_{3}(x,x_{p}^{\pm})\sigma_{3}(\tfrac{1}{x},x_{p}^{\pm}).
\]
Provided $\sigma_{3}(x)\sigma_{3}(\tfrac{1}{x})=1$ this is the inverse
of the crossing relation (5.44). 

For \eqref{eq:F-diff-light}, where the virtual particle is of the
opposite mass to the real particle, the relevant terms are $\mathsf{A}_{pq}^{LL'}$,
$\mathsf{B}_{pq}^{LL'}$, $\mathsf{A}_{pq}^{LR'}$, $\mathsf{C}_{pq}^{LR'}$,
and the phases are
\begin{align*}
S_{pq}^{LL'}=\sqrt{\frac{x_{p}^{-}}{x_{p}^{+}}}\sigma_{1},\qquad\tau_{pq}^{LR'} & =\frac{x-1/x_{p}^{-}}{x-1/x_{p}^{+}}\sqrt{\frac{x_{p}^{-}}{x_{p}^{+}}}\sigma_{1}.\\
S_{pq}^{LR'} & =\frac{1}{\zeta^{LR'}}\tau_{pq}^{LR'}=\sqrt{\frac{x-1/x_{p}^{-}}{x-1/x_{p}^{+}}}\sqrt{\frac{x_{p}^{-}}{x_{p}^{+}}}\sigma_{1}.
\end{align*}
Then we obtain the inverse of the crossing relation (5.46), as long
as  $\sigma_{1}(x)\sigma_{1}(\tfrac{1}{x})=1$: 
\[
S_{pq}^{LL'}S_{p\bar{q}}^{LR'}=\sqrt{\frac{x-x_{p}^{-}}{x-x_{p}^{+}}}\sqrt{\frac{x_{p}^{-}}{x_{p}^{+}}}\sigma_{1}(x)\sigma_{1}(\tfrac{1}{x}).
\]

\section{Algebraic Curves for \texorpdfstring{$AdS_{3}\times S^{3}$}{AdS3 x S3}
sans $T^{4}$\label{sec:Algebraic-Curves-sans-T4}}

This appendix sets up the ``$T^{4}$'' case in exactly the same
way as above, following \cite{Zarembo:2010yz}. The main reason for
doing so is in order to calculate F-term corrections, to illustrate
the limit $\alpha\to1$.  

The Cartan matrix is: 
\[
A=\left[\begin{array}{ccc}
0 & -1 & 0\\
-1 & 2 & -1\\
0 & -1 & 0
\end{array}\right]\otimes1_{2\times2}\,.
\]
Here is the basis given by \cite{Zarembo:2010yz}, with a minus inserted
on the left-hand half this time, and the order of the index $i$ chosen
so that the quasimomenta $q_{i}(x)$ match those in \cite{Gromov:2008ie,Gromov:2008ec}:
\newcommand{\mydynkinT}{
\rlap{\hspace{-2mm}
\begin{tikzpicture}[scale=0.55, darkred]
\draw (0,0) -- (3,0);
\draw [fill=white] (0,0) circle (2mm); 
\draw [fill=white] (1.5,0) circle (2mm); 
\draw [fill=white] (3,0) circle (2mm); 
\draw (-2mm,0) -- (2mm,0);
\draw (0,-2mm) -- (0,2mm);
\draw (28mm,0) -- (32mm,0);
\draw (3,-2mm) -- (3,2mm);
\end{tikzpicture}
}}
\[
\begin{array}{cccccccc|ccc}
H_{1} & H_{2} & H_{3} &  & H_{\bar{1}} & H_{\bar{2}} & H_{\bar{3}} &  &  &  & i\\
\mydynkinT &  &  &  & \mydynkinT &  &  & \\
-1 &  &  &  &  &  &  &  & F &  & \hat{1}\\
-1 & 1 &  &  &  &  &  &  & B &  & \tilde{1}\\
 & -1 & 1 &  &  &  &  &  & B &  & \tilde{4}\\
 &  & 1 &  &  &  &  &  & F & \vspace{2mm} & \hat{4}\\
 &  &  &  & 1 &  &  &  & F &  & \hat{2}\\
 &  &  &  & 1 & -1 &  &  & B &  & \tilde{2}\\
 &  &  &  &  & 1 & -1 &  & B &  & \tilde{3}\\
 &  &  &  &  &  & -1 &  & F &  & \hat{3}
\end{array}
\]
The inversion symmetry is the same as \eqref{eq:inv-cond-p} above,
or in terms of the $q_{i}$: 
\begin{align*}
\hat{q}_{1}(\tfrac{1}{x}) & =-\hat{q}_{2}(x),\qquad\hat{q}_{3}(\tfrac{1}{x})=-\hat{q}_{4}(x)\\
\tilde{q}_{1}(\tfrac{1}{x}) & =-\tilde{q}_{2}(x),\qquad\,\tilde{q}_{3}(\tfrac{1}{x})=-\tilde{q}_{4}(x).
\end{align*}
The vacuum is given by 
\begin{align*}
\kappa & =\frac{\Delta}{2g}\big(-1,0,-1,\:1,0,1\big)\\
q_{i}(x) & =\frac{\Delta}{2g}\frac{x}{x^{2}-1}\big(1,1,-1,-1,\;1,1,-1,-1\big).
\end{align*}

We can again make modes by colouring in, this time $2$ and $\bar{2}$
are the momentum-carrying nodes: \newcommand{\smalldynkinT}[3]{
\begin{tikzpicture}[scale=0.6]
\draw (1,0) -- (3,0);
\draw [fill=#1] (1,0) circle (2mm); 
\draw [fill=#2] (2,0) circle (2mm); 
\draw [fill=#3] (3,0) circle (2mm); 

\draw [thin] (8mm,0) -- (12mm,0);
\draw [thin] (1,-2mm) -- (1,2mm);

\draw [thin] (28mm,0) -- (32mm,0);
\draw [thin] (3,-2mm) -- (3,2mm);

\end{tikzpicture}
}  \begin{equation}  %
\begin{tabular}{lccc}
Bosons: & \smalldynkinT{white}{\colourone}{white} & $(\tilde{1},\tilde{4})$ & $(\tilde{2},\tilde{3})$\tabularnewline
 & \smalldynkinT{\colourone}{\colourone}{\colourone} & $(\hat{1},\hat{4})$ & $(\hat{2},\hat{3})$\tabularnewline
Fermions: & \smalldynkinT{\colourone}{\colourone}{white} & $(\hat{1},\tilde{4})$ & $(\hat{2},\tilde{3})$\tabularnewline
 & \smalldynkinT{white}{\colourone}{\colourone} & $(\tilde{1},\hat{4})$ & $(\tilde{2},\hat{3})$\tabularnewline
\end{tabular} \label{eq:T4-modes-colouring-in} \end{equation}  where again $\smallcirc{\colourone}=1$
on the left but $\smallcirc{\colourone}=-1$ on the right. Then the
asymptotic behaviour of the modes is \vspace{-2mm}

\begin{equation}
\left(\begin{array}{c}
\delta\hat{q}_{1}\\
\delta\hat{q}_{2}\\
\delta\hat{q}_{3}\\
\delta\hat{q}_{4}\\
\hline \delta\tilde{q}_{1}\\
\delta\tilde{q}_{2}\\
\delta\tilde{q}_{3}\\
\delta\tilde{q}_{4}
\end{array}\right)\to\frac{1}{2gx}\begin{pmatrix}\begin{array}{r}
\hphantom{{+}}\delta\Delta+N_{\hat{1}\hat{4}}+N_{\hat{1}\tilde{4}}\\
\hphantom{{+}}\delta\Delta+N_{\hat{2}\hat{3}}+N_{\hat{2}\tilde{3}}\\
-\delta\Delta-N_{\hat{2}\hat{3}}-N_{\tilde{2}\hat{3}}\\
-\delta\Delta-N_{\hat{1}\hat{4}}-N_{\tilde{1}\hat{4}}\\
\hline -N_{\tilde{1}\tilde{4}}\quad-N_{\tilde{1}\hat{4}}\\
-N_{\tilde{2}\tilde{3}}\quad-N_{\tilde{2}\hat{3}}\\
+N_{\tilde{2}\tilde{3}}\quad+N_{\hat{2}\tilde{3}}\\
+N_{\tilde{1}\tilde{4}}\quad-N_{\hat{1}\tilde{4}}
\end{array}\end{pmatrix}+\ldots\label{eq:T4-asympt-Nij}
\end{equation}
which matches the 1st \& 4th columns of \cite{Gromov:2008ec}'s (A.9),
apart from normalisation of $\Delta$. As expected this is very much
like $AdS_{5}\times S^{5}$ with the modes connecting left and right
turned off. Comparing this with \eqref{eq:dq-N-infinity} at $\phi=0$,
the bosons match up perfectly but for the fermions things aren't so
simple.

\subsection{F-term Corrections}

Using a giant magnon $G_{2}(x)=G_{\mathrm{mag}}(x)$, i.e. a giant
$(\tilde{1},\tilde{4})$ mode, here are some of the integrands $F^{(\ell)}(x)=\sum_{ij}(-1)^{F_{ij}}\exp(-i\ell(q_{i}-q_{j}))$,
showing terms in the same order as \eqref{eq:T4-modes-colouring-in}
above: 
\begin{align}
F_{\mathrm{left}}^{(1)} & =e^{-\frac{\Delta}{g}\frac{i\, x}{x^{2}-1}}\left[\frac{\Xp}{\Xm}\left(\frac{x-\Xm}{x-\Xp}\right)^{2}+1-\frac{x-\Xm}{x-\Xp}\sqrt{\frac{\Xp}{\Xm}}-\frac{x-\Xm}{x-\Xp}\sqrt{\frac{\Xp}{\Xm}}\right]\displaybreak[0]\label{eq:T4-F-left}\\
F_{\mathrm{right}}^{(1)} & =e^{-\frac{\Delta}{g}\frac{i\, x}{x^{2}-1}}\left[\frac{\Xm}{\Xp}\left(\frac{1-x\Xp}{1-x\Xm}\right)^{2}+1-\frac{1-x\Xp}{1-x\Xm}\sqrt{\frac{\Xm}{\Xp}}-\frac{1-x\Xp}{1-x\Xm}\sqrt{\frac{\Xm}{\Xp}}\right]\nonumber 
\end{align}
These line up with the $AdS_{3}\times S^{3}\times S^{3}$ ones above,
in total 
\[
F_{m_{3}}^{(1)}\Big\vert_{\phi=\pi/2}+F_{m_{4}}^{(1)}=F_{\mathrm{left}}^{(1)}+F_{\mathrm{right}}^{(1)}
\]
but also term by term.

\section{Worldsheet Theory\label{sec:Worldsheet-Giant-Magnons}}

The bosonic action in conformal gauge (and setting $\alpha'=1$) is
\[
\mathcal{S}=\int\frac{d\tau d\sigma}{4\pi}\left(R^{2}\partial_{\mu}\bar{Z}\cdot\partial^{\mu}Z+\frac{R^{2}}{\cos^{2}\phi}\partial_{\mu}\bar{X}\cdot\partial^{\mu}X+\frac{R^{2}}{\sin^{2}\phi}\partial_{\mu}\bar{Y}\cdot\partial^{\mu}Y+R^{2}\partial_{\mu}\psi\partial^{\mu}\psi\right)
\]
where $\left|Z\right|^{2}=-1$ describes $AdS_{3}$ embedded in $\mathbb{C}^{1,1}$,
and $\left|X\right|^{2}=\left|Y\right|^{2}=1$ describe the two spheres
each in $\mathbb{C}^{2}$. The equations of motion are 
\begin{align*}
0 & =\partial_{\mu}\partial^{\mu}Z+\big(\partial_{\mu}\bar{Z}\cdot\partial^{\mu}Z\big)Z & 0 & =\partial_{\mu}\partial^{\mu}Y+\big(\partial_{\mu}\bar{Y}\cdot\partial^{\mu}Y\big)Y\\
0 & =\partial_{\mu}\partial^{\mu}X+\big(\partial_{\mu}\bar{X}\cdot\partial^{\mu}X\big)X & 0 & =\partial_{\mu}\partial^{\mu}\psi
\end{align*}
with the four components coupled only through the Virasoro constraints
\begin{align*}
0 & =\partial_{\tau}\bar{Z}\cdot\partial_{\tau}Z+\partial_{\sigma}\bar{Z}\cdot\partial_{\sigma}Z+\smash{\frac{1}{\cos^{2}\phi}}\big(\partial_{\tau}\bar{X}\cdot\partial_{\tau}X+\partial_{\sigma}\bar{X}\cdot\partial_{\sigma}X\big)\\
 & \qquad\qquad\qquad\qquad+\frac{1}{\sin^{2}\phi}\big(\partial_{\tau}\bar{Y}\cdot\partial_{\tau}Y+\partial_{\sigma}\bar{Y}\cdot\partial_{\sigma}Y\big)+\big(\partial_{\tau}\psi\:\partial_{\tau}\psi+\partial_{\sigma}\psi\:\partial_{\sigma}\psi\big)\displaybreak[0]\\
0 & =\partial_{\tau}\bar{Z}\cdot\partial_{\sigma}Z+\frac{1}{\cos^{2}\phi}\big(\partial_{\tau}\bar{X}\cdot\partial_{\sigma}X\big)+\frac{1}{\sin^{2}\phi}\big(\partial_{\tau}\bar{Y}\cdot\partial_{\sigma}Y\big)+\mbox{c.c.}\quad+2\partial_{\tau}\psi\:\partial_{\sigma}\psi.
\end{align*}
The global charges are 
\begin{align*}
\Delta & =R^{2}\int_{-L}^{L}\frac{d\sigma}{2\pi}\im(\bar{Z}_{0}\:\partial_{\tau}Z_{0})\\
J_{X} & =\frac{R^{2}}{\cos^{2}\phi}\int_{-L}^{L}\frac{d\sigma}{2\pi}\im(\bar{X}_{1}\:\partial_{\tau}X_{1}),\qquad J_{Y}=\frac{R^{2}}{\sin^{2}\phi}\int_{-L}^{L}\frac{d\sigma}{2\pi}\im(\bar{Y}_{1}\:\partial_{\tau}Y_{1})\negthickspace\negthickspace\negthickspace\negthickspace\negthickspace\displaybreak[0]\\
J' & =\cos^{2}\phi\: J_{X}+\sin^{2}\phi\: J_{Y}\\
\shortintertext{and}P & =\int d\sigma\im(\partial_{\sigma}\log X_{1})+\int d\sigma\im(\partial_{\sigma}\log Y_{1}).
\end{align*}

The spinning string solution studied by \cite{Forini:2012bb} is
\begin{equation}
Z_{0}=e^{i\kappa\tau}\cosh\rho(\sigma),\qquad Z_{1}=e^{i\omega\tau}\sinh\rho(\sigma),\qquad X_{1}=e^{i\nu_{+}\tau},\qquad Y_{1}=e^{i\nu_{-}\tau}\label{eq:WS-spinning-string}
\end{equation}
for which the Virasoro constraint gives 
\[
0=-\kappa^{2}\cosh^{2}\rho+\omega^{2}\sinh^{2}\rho+\rho^{\prime2}+\frac{1}{\cos^{2}\phi}\nu_{+}^{2}+\frac{1}{\sin^{2}\phi}\nu_{-}^{2}\,.
\]
Here we consider only the case $\nu_{+}=\cos^{2}\phi\:\nu$, $\nu_{-}=\sin^{2}\phi\:\nu$,
giving $J_{X}=J_{Y}=\nu LR^{2}/2\pi$. This reduces to the supersymmetric
BMN point particle when $\rho=0$, and $\kappa=\nu$ \cite{Berenstein:2002jq},
see also \cite{Russo:2002rq,Lu:2002kw,Gomis:2002qi,Gava:2002xb,Sommovigo:2003kd}
for this background. At $\phi=0$ this is stationary on the $Y$ sphere,
and the last term drops out of the Virasoro constraint. 

Now consider placing magnons into this space. Treating immediately
the finite-$J$ case, let $X_{\mathrm{fin}}(\sigma,\tau)$ be a solution
in $\mathbb{R}\times S^{3}$ in conformal gauge and with $t=\tau$:
it satisfies $\partial_{\tau}\bar{X}\cdot\partial_{\tau}X+\partial_{\sigma}\bar{X}\cdot\partial_{\sigma}X=1$.
Let $2L_{\mathrm{fin}}$ be the periodicity in $\sigma$ (i.e. the
distance between two cusps). Writing charges using this as $\Delta_{\mathrm{fin}}$
(and $J_{\mathrm{fin}}=J'\vert_{\phi=0}$) it has dispersion relation
\[
\Delta_{\mathrm{fin}}-J_{\mathrm{fin}}=4g\sin\frac{p}{2}\Big(1-4\,\sin^{2}\frac{p}{2}\: e^{-2\Delta_{\mathrm{fin}}\big/4g\sin\frac{p}{2}}+\ldots\Big).
\]
The solution at general $\phi$ is 
\begin{align}
X(\sigma,\tau) & =X_{\mathrm{fin}}(\cos^{2}\phi\:\sigma,\cos^{2}\phi\:\tau)\label{eq:X-embed}\\
Y_{1}(\sigma,\tau) & =e^{i\sin^{2}\phi\:\tau},\qquad Y_{2}=0\nonumber 
\end{align}
with $\cos^{2}\phi\: L=L_{\mathrm{fin}}$, thus $\Delta=\tfrac{1}{\cos^{2}\phi}\Delta_{\mathrm{fin}}$.
This has $P=p$ and 
\begin{equation}
\Delta-J'=\Delta_{\mathrm{fin}}-J_{\mathrm{fin}}=4g\sin\frac{p}{2}\Big(1-4\,\sin^{2}\frac{p}{2}\: e^{-2\Delta\cos^{2}\phi\big/4g\sin\frac{p}{2}}+\ldots\Big).\label{eq:dE-mu-WS}
\end{equation}
The exponent in the finite-$J$ correction is exactly what we saw
in the algebraic curve calculation \eqref{eq:dE-mu-final}, apart
from here considering a magnon in the other sphere i.e. a giant ``3''
mode. 

Other solutions can be similarly embedded, in particular:
\begin{itemize}
\item We can use the same scattering solutions as usual, \cite{Spradlin:2006wk,Kalousios:2006xy,Kalousios:2010ne},
within one sphere. The time delay for scattering solutions is defined
like this (initially on a unit sphere):
\[
X_{\mathrm{scat}}(\sigma,\tau)=\begin{cases}
X_{\mathrm{mag}}(\sigma,\tau), & \sigma,\tau\to-\infty\\
X_{\mathrm{mag}}(\sigma,\tau-\Delta\tau_{\mathrm{mag}}), & \sigma,\tau\to+\infty.
\end{cases}
\]
It is clear that embedding this solution into the $S_{+}^{3}$ sphere
via \eqref{eq:X-embed} will give us a time delay $\Delta\tau=\frac{1}{\cos^{2}\phi}\tan\frac{p}{2}\log(\cos^{2}\frac{p}{2})$
scaled from the usual (centre of mass frame) delay \cite{Hofman:2006xt}.
\item We can embed a different giant magnon into each sphere, and they don't
talk to each other at all. Thus we would expect the relevant terms
in the S-matrix to be 1, and this is exactly what we saw in \eqref{eq:F-diff-light}
above: the trivial entries there correspond to a physical 3 mode and
a virtual 1 or $\bar{1}$.
\end{itemize}

\bibliographystyle{my-JHEP-4}
\bibliography{/Users/me/Documents/Papers/complete-library-processed,complete-library-processed}

\end{document}